\documentclass[aps,prx,showpacs,twocolumn,reprint,superscriptaddress]{revtex4-1}

\usepackage[dvips]{graphicx}
\usepackage{amsmath,amssymb,amsthm,mathrsfs,amsfonts,dsfont,mathtools}
\usepackage[unicode=true, colorlinks=true]{hyperref}
\usepackage{blkarray}
\usepackage{epsfig}
\usepackage{braket}
\usepackage{bm}
\usepackage{enumerate}
\usepackage{color}
\usepackage{graphicx}
\usepackage{pgf}
\usepackage{pgfplots}
\pgfplotsset{compat=1.17}
\usepackage{tikz}
\usetikzlibrary{quantikz}
\usepackage{enumitem}
\usepackage[capitalize]{cleveref}
\usepackage[normalem]{ulem}
\usepackage{amsthm}
\usepackage{physics}

\usepackage{xcolor}

\newtheorem{statement}{Statement}

\Crefname{theorem}{Theorem}{Theorems}

\begin{document}
\title{Simulating non-unitary dynamics using quantum signal processing with unitary block encoding}

\author{Hans Hon Sang Chan}

    \affiliation{Quantinuum, 13-15 Hills Road, CB2 1NL Cambridge, United Kingdom}
    \affiliation{Department of Materials, University of Oxford, Parks Road, Oxford OX1 3PH, United Kingdom}

\author{David Mu\~noz Ramo}
\affiliation{Quantinuum, 13-15 Hills Road, CB2 1NL Cambridge, United Kingdom}

\author{Nathan Fitzpatrick}
    \email{nathan.fitzpatrick@quantinuum.com}% Your name
\affiliation{Quantinuum, 13-15 Hills Road, CB2 1NL Cambridge, United Kingdom}

\date{\today} % Leave empty to omit a date

\begin{abstract}
\noindent
We adapt a recent advance in resource-frugal quantum signal processing -- the Quantum Eigenvalue Transform with Unitary matrices (QET-U) -- to explore non-unitary imaginary time evolution on early fault-tolerant quantum computers using exactly emulated quantum circuits. We test strategies for optimising the circuit depth and the probability of successfully preparing the desired imaginary-time evolved states. For the task of ground state preparation, we confirm that the probability of successful post-selection is quadratic in the initial reference state overlap $\gamma$ as $O(\gamma^2)$. When applied instead to thermal state preparation, we show QET-U can directly estimate partition functions at exponential cost. Finally, we combine QET-U with Trotter product formula to perform non-normal Hamiltonian simulation in the propagation of Lindbladian open quantum system dynamics. We find that QET-U for non-unitary dynamics is flexible, intuitive and straightforward to use, and suggest ways for delivering quantum advantage in simulation tasks.
\end{abstract}

\maketitle

\section{Introduction} \label{sec:introduction}
Non-unitary imaginary-time evolution (ITE) is a powerful and versatile technique in quantum many-body simulations, often discussed in the context of preparing the lowest energy eigenstate of a Hamiltonian matrix.
It is useful in other simulation tasks, for example, modelling quantum states that are entangled to another as part of a subsystem. This has applications in predicting finite-temperature statistical mechanics as well as the non-unitary dynamics of open quantum systems.

The time and memory cost of performing ITE on a classical computer grows exponentially with system size. In recent years, many quantum computing algorithms for approximate-ITE have been proposed which promise to overcome these prohibitive scaling costs. These methods use different techniques to circumvent the fact that the ITE operator is non-unitary and cannot be directly implemented using quantum circuit gate operations alone: the four main classes of quantum imaginary time algorithms are  variational approaches (VITE)~\cite{mcardle2019variational,harriet2020, Marcello2021, fitzite, Matsumoto_2022, Sokolov2022}, tensor network compiled circuits for imaginary time evolution~\cite{lubash}, the quantum imaginary time evolution (QITE) method~\cite{motta2020determining}, the probabilistic imaginary time evolution (PITE)~\cite{Kosugi2022}, and quantum signal processing (QSP)~\cite{Coopmans2022, Silva2021}. The key metrics in comparing the utility of these methods, beyond circuit complexity, are the probability of successfully preparing the desired state and the necessary classical pre- and post-processing resource.
% \begin{itemize}
%     \item Variational imaginary time evolution (VITE)~\cite{mcardle2019variational, Matsumoto_2022, Sokolov2022}, where a parameterised quantum circuit ansatz is classically optimised to approximate ITE;
%     \item Quantum imaginary time evolution (QITE)~\cite{motta2020determining}, which uses Pauli tomography to approximate non-unitary imaginary-time propagation with Trotterized unitary evolution;
%     \item Probabilistic imaginary time evolution (PITE)~\cite{Kosugi2022, Chan2022}, which approximates a small imaginary time step by entangling real-time Hamiltonian evolution to an ancilla qubit and partially collapsing the state by performing a mid-circuit or weak measurements on the ancilla, attaining long imaginary time evolution by repeated application;
%     \item Quantum signal processing (QSP), which approximates the ITE operator with a polynomial transform of the Hamiltonian matrix~\cite{Coopmans2022, Silva2021}.
% \end{itemize}

QSP stands out among these methods. Apart from providing a framework for unifying all quantum algorithms with known speedups~\cite{Chuang2021}, it has been observed to be the optimal approach in the long term for simulating real-time Hamiltonian evolution on quantum computers~\cite{low2017optimal}. This is perhaps not surprising; the technique of expanding both real- and imaginary-time evolution operators in a basis of Chebyshev polynomial transforms, which is the central tenet in QSP methods, has been known since as early as the 90's to provide some of the most accurate quantum wavepacket dynamics simulations (see e.g. the Chebyshev scheme~\cite{Leforestier1991}) in chemical physics. However, the quantum resource cost for performing QSP is substantial. For example, using Linear-Combination of Unitaries (LCU) subroutine to encode the Hamiltonian matrix~\cite{Chuang2021} requires an entire register of error-corrected ancilla qubits with a size proportional to log of the number of expansion terms in the Hamiltonian and many associated entangling gates. This places the prospect of performing QSP well into the mature fault-tolerant quantum computing regime.

Recently, there have been significant advances in `early fault-tolerant' quantum algorithms which are expected to use fewer error-corrected qubits~\cite{Tong2022designingalgorithms} and practical error detection protocols~\cite{Marcello2022}. In particular, QSP algorithms which use few or even only a single additional ancilla qubit~\cite{dong2022ground, McArdle2022, Silva2021, Silva2022, Wan_2022} have been proposed. These techniques circumvent the expensive LCU block-encoding model, and opt instead to use e.g. ancilla-controlled unitary Hamiltonian simulation (or the real-time evolution oracle) as the Hamiltonian input model. Using controlled unitary evolution this way is employed in many other early fault-tolerant state preparation algorithms, from standard routines like the Hadamard test and Kitaev's iterative phase estimation to the PITE approach~\cite{Kosugi2022} and the rodeo algorithm~\cite{PhysRevLett.127.040505, Meister2022}.

In this work, we perform imaginary-time propagation on emulated quantum circuits using Quantum Eigenvalue Transform with Unitary block-encoding (QET-U)~\cite{dong2022ground, Silva2021}, a recently-introduced early fault-tolerant QSP algorithms which leverages stochastic measurement of only a single ancilla qubit and efficient QSP phase factor evaluation. We consolidate the similarities between the basic quantum circuit `\textit{motifs}' this approach shares with other state preparation algorithms to apply the Wick-rotated Schr\"odinger equation using QET-U exactly (up to the polynomial approximation).
This general approach, similarly explored previously in e.g.~\cite{Silva2021}, presents many advantages over the other quantum algorithms for performing ITE; it avoids cumbersome classical computational steps such as variational optimisation (necessary in VITE) and Pauli tomography (necessary in QITE). Unlike PITE, QET-U can be implemented such that it does not require sequential successful stochastic ancilla measurements. However, like PITE and QITE, the query complexity  of simulating imaginary-time of $\tau$ with error $\epsilon$ is expected to scale linearly with the imaginary time as $\propto O(\tau\log\epsilon^{-1})$~\cite{Silva2021}, and as we will see presently in the case of QET-U the corresponding circuits are far too deep for near-term hardware.  Furthermore, the original implementation of QET-U requires bounds of the extremum Hamiltonian eigenvalues. These are well-understood challenging tasks, and we point to some possible solutions herein. 
% , nor does it require a large qubit overhead for performing LCU block-encoding. %sHowever, the polynomial 

We apply our techniques to numerical qubit emulations, targeting three key simulation tasks: ground state preparation, thermal state preparation, and, for the first time, demonstrate quantum simulation of open quantum system dynamics using QSP. We conclude with a discussion on how techniques explored here may help us exploit early fault-tolerant quantum computers sooner for quantum simulation and applications beyond.

This manuscript is structured as follows. In \cref{sec:background}, we summarise the machinery of QSP and QET-U. In \cref{sec:methods}, we describe our specific approach of using QET-U for performing non-unitary ITE, and contextualise our application of this technique in preparing ground state and computing thermal state properties. In \cref{sec:results}, we report small proof-of-concept numerical simulations of these use cases and discuss our findings. We finish with concluding thoughts in \cref{sec:conclusion}.

\begin{figure*}
    % \begin{subfigure}{.7\textwidth}
    % \centering
    % \caption{Single qubit QSP}\label{fig:singleQSP}
    \begin{tikzpicture}
    \node [scale=0.7]{
    \begin{quantikz}[row sep={15pt,between origins}, column sep=5pt]
    &\lstick{$\ket{0}$} & \gate[style={fill=pink}]{R_x(-\varphi_0)} & \gate[style={fill=green!20}]{R_z(\theta t+\sigma)} & \gate[style={fill=red!20}]{R_x(-\varphi_1)} & \gate[style={fill=green!20}]{R_z(-(\theta t+\sigma))} & \ \ldots\ \qw & \gate[style={fill=red!20}]{R_x(-\varphi_1)} & \gate[style={fill=green!20}]{R_z(-(\theta t+\sigma))} & \gate[style={fill=red!20}]{R_x(-\varphi_0)} &\meterD{0} 
    \end{quantikz}
    };
    \node at (0, -1)[scale=0.7]{$
    \textcolor{red!80}{\begin{bmatrix} 
    \cos\varphi_0 & i\sin\varphi_0 \\
    i\sin\varphi_0 & \cos\varphi_0 \\
    \end{bmatrix}}
    \textcolor{green!90}{\begin{bmatrix} 
    e^{i(\theta t+\sigma)} & 0 \\
    0 & e^{-i(\theta t+\sigma)} \\
    \end{bmatrix}}
    \textcolor{red!80}{\begin{bmatrix} 
    \cos\varphi_1 & i\sin\varphi_1 \\
    i\sin\varphi_1 & \cos\varphi_1 \\
    \end{bmatrix}}
    \textcolor{green!90}{\begin{bmatrix} 
    e^{-i(\theta t+\sigma)} & 0 \\
    0 & e^{i(\theta t+\sigma)} \\
    \end{bmatrix}}\dots
    = 
    \begin{blockarray}{ccc}
    \begin{block}{[cc]c}
      \sum_n A_n\{\varphi_i\}\cos(2n(\theta t + \sigma)) & * & \ket{0} \\
      * & * & \ket{1} \\
    \end{block}
    \end{blockarray}
    $};
    % \end{tikzpicture}
    % \end{subfigure}%
    % \begin{subfigure}{.3\linewidth}
    %     \centering
    %     \caption{PITE circuit}\label{fig:PITE}
        % \begin{tikzpicture}
        
    \node at (9, -0.5) [scale=0.75]{
        \begin{quantikz}[row sep={28pt,between origins}, column sep=5pt]
        & \lstick{$\ket{0}$} & \gate[style={fill=pink}]{H} & \gate[style={fill=green!20}]{R_z(\sigma)} & \octrl{1} & \ctrl{1} & \gate[style={fill=pink}]{H} & \meterD{0} \\
        & &\lstick[wires=1]{$\ket{\psi_0} \{$} &\qwbundle[alternate]{}&\gate[style={fill=green!20}]{e^{i\mathcal{H}t}}\qwbundle[alternate]{} &\gate[style={fill=green!20}]{e^{-i\mathcal{H}t}}\qwbundle[alternate]{}& \qwbundle[alternate]{}&
        \end{quantikz}
        };
        % \end{tikzpicture}
        % \hfill
    % \end{subfigure}%
    % \vspace{0.5cm}
    % \begin{subfigure}[b]{\linewidth}
    % \caption{QET-U with forward and reverse time evolution}\label{fig:qetu_forrev}
    % \begin{tikzpicture}
    \node at (3, -2.6) [scale=0.8]{
    \begin{quantikz}[row sep={28pt,between origins}, column sep=5pt]
    &\lstick{$\ket{0}$} & \gate[style={fill=pink}]{R_x(-\varphi_0)} & \gate[style={fill=green!20}]{R_z(\sigma)} & \octrl{1} & \ctrl{1} & \gate[style={fill=red!20}]{R_x(-\varphi_1)} & \gate[style={fill=green!20}]{R_z(-\sigma)} & \octrl{1} & \ctrl{1} &  \ \ldots\ \qw & \gate[style={fill=green!20}]{R_z(-\sigma)} & \octrl{1} & \ctrl{1} & \gate[style={fill=red!20}]{R_x(-\varphi_0)} &\meterD{0} \\
    &\lstick[wires=1]{$\ket{\psi_0} \{$} &\qwbundle[alternate]{}&\qwbundle[alternate]{}&\gate[style={fill=green!20}]{e^{i\mathcal{H}t}}\qwbundle[alternate]{} &\gate[style={fill=green!20}]{e^{-i\mathcal{H}t}}\qwbundle[alternate]{}& \qwbundle[alternate]{}&\qwbundle[alternate]{}&\gate[style={fill=green!20}]{e^{-i\mathcal{H}t}}\qwbundle[alternate]{} &\gate[style={fill=green!20}]{e^{i\mathcal{H}t}}\qwbundle[alternate]{}  & \ \ldots\ \qwbundle[alternate]{} & \qwbundle[alternate]{} &\gate[style={fill=green!20}]{e^{-i\mathcal{H}t}}\qwbundle[alternate]{} &\gate[style={fill=green!20}]{e^{i\mathcal{H}t}}\qwbundle[alternate]{} &  \qwbundle[alternate]{}\rstick{$\frac{1}{\sqrt{\mathcal{N}}}P(\mathcal{H})\ket{\psi_0}$}
    \end{quantikz}
    };
    \node at (3, -4.0)[scale=0.8]{$
    \textcolor{red!80}{\begin{bmatrix} 
    \cos\varphi_0\hat{I} & i\sin\varphi_0\hat{I} \\
    i\sin\varphi_0\hat{I} & \cos\varphi_0\hat{I} \\
    \end{bmatrix}}
    \textcolor{green!90}{\begin{bmatrix} 
    e^{i(\mathcal{H} t+\sigma)} & 0 \\
    0 & e^{-i(\mathcal{H} t+\sigma)} \\
    \end{bmatrix}}
    \textcolor{red!80}{\begin{bmatrix} 
    \cos\varphi_1\hat{I} & i\sin\varphi_1\hat{I} \\
    i\sin\varphi_1\hat{I} & \cos\varphi_1\hat{I} \\
    \end{bmatrix}}
    \textcolor{green!90}
    {\begin{bmatrix} 
    e^{-i(\mathcal{H}t+\sigma)} & 0 \\
    0 & e^{i(\mathcal{H} t+\sigma)} \\
    \end{bmatrix}}\dots
    =
    \begin{blockarray}{ccc}
    \begin{block}{[cc]c}
      \sum_n A_n\{\varphi_i\}\cos(2n(\mathcal{H} t + \sigma)) & * & \ket{\psi_0}\ket{0} \\
      * & * & \ket{\psi_0}\ket{1} \\
    \end{block}
    \end{blockarray}
    $};
    \end{tikzpicture}
    % \end{subfigure}
    
    \caption{Circuit for QET-U and relation to the PITE circuit. It can be seen that the operations of the single qubit rotation sequence (TOP LEFT) and the QET-U circuit (BELOW) manifest identical transform, except the former performs an eigenvalue transform on the variable $\theta$ and the latter a transform on the matrix $\mathcal{H}$. The PITE approach (TOP RIGHT), performs a $\propto\cos(I +\mathcal{H}t)$ transform to approximate imaginary-time evolution up to first order.}
    \label{fig:QETU}
\end{figure*}
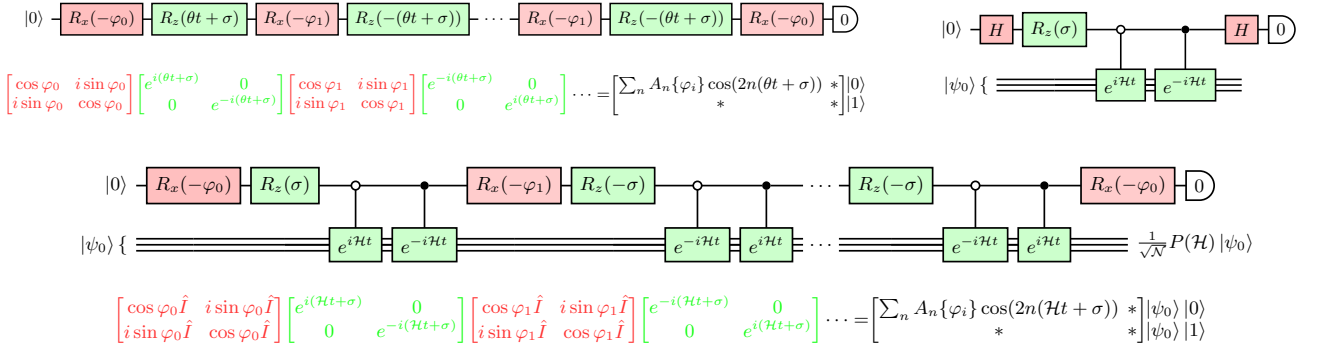

\section{Background} \label{sec:background}
\subsection{Overview of QSP and QET-U}
We review the components of QSP and QET-U relevant to this work here, but provide a more detailed account in the Appendix. The expert reader may skip to \cref{sec:methods}.

Quantum Signal Processing is a class of methods for preparing matrix functions on a quantum computer. Consider a sparse $n$-qubit ($2^n\times 2^n$) Hamiltonian $\mathcal{H}=\sum_k a_k\mathcal{P}_k$ where $\mathcal{P}_k$ are a polynomial number of Pauli operators. Given an $n$-qubit initial state $\ket{\psi_0}$ and an input model for embedding $\mathcal{H}$, QSP manifests a degree $d$ matrix polynomial transformation $P$,
\begin{equation} \label{eq:QSP transform}
\ket{\psi_0} \rightarrow P(\mathcal{H})\ket{\psi_0}
\end{equation}
up to some normalisation. It achieves this with a repeating circuit structure (a \textit{QSP sequence}) that interleaves a fixed \textit{signal operator} -- the aforementioned Hamiltonian input model that embeds $\mathcal{H}$
into the top left block of a larger unitary matrix $W_x(\mathcal{H})$, with a sequence of \textit{signal processing operators} -- single-qubit rotations with classically pre-computed rotation angles or \textit{phase factors} $\{\varphi_k\}$ that act on the embedded $\mathcal{H}$ and are responsible for realising the desired $P(\mathcal{H})$.   
It can be shown that the polynomial transform is actually applied onto the eigenvalues of $\mathcal{H}$ that have been encoded on a single ancilla, hence the alternative name for this machinery the Quantum Eigenvalue Transform for when $\mathcal{H}$ is Hermitian. The unitary of the QSP sequence is thus:
\begin{align*}
    U_x(\mathcal{H}) = e^{i\varphi_0 Z} &\prod_{k=1}^d W_x(\mathcal{H}) e^{i\varphi_k Z} = \left[ \begin{array}{cc} P(\mathcal{H}) & * \\ * & *  \end{array} \right ], \\ 
    \text{where }W_x(\mathcal{H}) &=\left[ \begin{array}{cc} \mathcal{H} & i\sqrt{1-\mathcal{H}^2} \\ i\sqrt{1-\mathcal{H}^2} & \mathcal{H}  \end{array} \right].
\end{align*}
Upon application of the QSP sequence circuit to $\ket{\psi_0}$, the signal processing ancilla qubit is measured in the computational basis. If it is in the $\ket{0}$ state, the post-selected $n$-qubit state should correspond to achieving the transformation of \cref{eq:QSP transform}. The versatility of QSP boils down to the fact that any analytic function of the Hamiltonian $f(\mathcal{H})$ can be arbitrarily well-approximated on a quantum computer, through the classical optimisation of phase factors $\{\varphi_k\}$ that generates a polynomial approximation of the transform $P(\mathcal{H})\approx f(\mathcal{H})$. The query complexity for performing sparse Hamiltonian simulation using QSP was found to be optimal~\cite{low2017optimal}.

In practice, additional non-unitary quantum operations are used to implement block encoding of the Hamiltonian e.g. LCU, which uses extra ancilla qubits that increase logarithmically with the number of terms in the Hamiltonian. The desired transformation of the Hamiltonian eigenvalues is thus only achieved if every ancilla qubit for LCU are measured to be in the $\ket{0}$ state -- incurring an associated measurement (or shots) overhead.

As we have also seen, the number of phase factors in the QSP sequence is linear in $d$; in general, high order approximations, thus the deep circuits, are necessary to perform the desired transform with high accuracy. Thanks to recent innovations, such as using a symmetric set of phase factors which we explore here, the cost for classical pre-computation of phase factors for high-degree polynomials is no longer an inhibiting step.

The recently-proposed QET with Unitary matrices (QET-U) approach is able to apply the desired polynomial transform of a Hermitian Hamiltonian matrix using only a single extra ancilla qubit~\cite{dong2022ground}. It uses the alternative $W_z$-convention of the QSP sequence, where the signal operator is diagonal, and instead $R_x$ rotations are performed on the ancilla qubit:
\begin{equation}
    U_z(\mathcal{H}) = e^{i\varphi_0 X} \prod_{k=1}^d W_z(\mathcal{H}) e^{i\varphi_k X}.
\end{equation}
Replacing the block-encoding signal operator with the forwards-backwards real-time evolution oracle $e^{\pm i\mathcal{H}t}$~\cite{Kosugi2022}:
\begin{equation}
    W_z(\mathcal{H}) = \left[ \begin{array}{cc} e^{-i\mathcal{H}t} & 0 \\ 0 & e^{i\mathcal{H}t}  \end{array} \right ]
\end{equation}
enables the embedding of Hamiltonian eigenvalues for QSP simply by performing Trotter evolution; $t$ is divided into regular time steps $t=N\delta t$ such that $e^{-i\mathcal{H}t} = (e^{-i\mathcal{H}\delta t})^N$, and each time step propagator is then partitioned into a product of propagators 
\begin{equation}
    e^{-i\mathcal{H}\delta t} = \prod_k e^{-ia_k\mathcal{P}_k\delta t} + \mathcal{O}(\delta t^2),
\end{equation}
for every term in the Hamiltonian decomposition -- see the Appendix. This avoids the additional qubit overhead for performing LCU block-encoding.
Specifically, the transformation achieved after application of the QSP sequence is:
\begin{equation} \label{eq:qet-u}
    \ket{\psi_0} \rightarrow \frac{P(\mathcal{H})}{\sqrt{\bra{\psi_0}P(\mathcal{H})^2\ket{\psi_0}}}\ket{\psi_0} = \frac{1}{\sqrt{\mathcal{N}}} P(\mathcal{H})\ket{\psi_0},
\end{equation}
where $1/\sqrt{\mathcal{N}}$ is a normalisation factor and $\mathcal{N}$ is the probability of success. The Hamiltonian function
\begin{equation}
    P(\mathcal{H}) = \sum^{d/2}_{k=0}c_kT_{2k}(\cos \mathcal{H}) 
\end{equation}
is a $d$-degree polynomial, which is a superposition of Chebyshev polynomials of $\cos \mathcal{H}
$, and $P^\dagger(\mathcal{H})=P(\mathcal{H})$.

The unitary nature of the real-time evolution oracle does imply that eigenvalue transform performed with QET-U is limited to Hermitian input matrices. Using Trotterisation to implement the time evolution oracle, while resource-friendly, introduces an additional Trotter error atop the polynomial approximation of the target function, but can be controlled with a combination of fine time-stepping and high-order product formulae. 
Moreover, the probability of successful ancilla measurement $\mathcal{N}$ is sensitive to the initial state $\ket{\psi_0}$ and the desired transform. Post-selection implies that a low probability of success can incur a high shots overhead.

Finally, QSP requires that the embedded Hamiltonian is scaled by its operator norm; e.g. in LCU,  every Hamiltonian term is normalised by the operator norm. In QET-U, this manifests as the restriction that any eigenvalue transform $P(\mathcal{H})$ must have a range $[-1, 1]$, for a Hamiltonian with eigenvalues between the domain $[\eta, \frac{\pi}{2}-\eta]$, where $\eta$ is chosen to be a small constant for improved optimisation of the phase factors. As a result, one must scale and shift the energy spectrum of the Hamiltonian $\mathcal{H}^\text{scale}$ such that:
\begin{align}
    E^\text{scale}_0&=\eta \label{eq:Escale_0}\\
    E^\text{scale}_{N-1}&=\frac{\pi}{2}-\eta. \label{eq:Escale_N}   
\end{align}
Optionally, one might also scale the range of $f(\mathcal{H})$.

The authors of QET-U propose pre-computing (at least) the lower bounds of the ground state energy $\Tilde{E}_0$ and maximum energy eigenvalue $\Tilde{E}_{N-1}$, and using those to embed $\mathcal{H}^\text{scale}$. This is achieved with an additional global phase $\sigma$ and scaling of the evolution time $t$ of the time evolution oracle:
\begin{align}
    e^{-i\mathcal{H}^\text{scale}} &= e^{-i(\mathcal{H}t + \sigma)}, \label{eq: scaled Ham}\quad \text{where}\\
    t &= \frac{\pi-\eta}{\Tilde{E}_{N-1}-\Tilde{E}_0}\\
    \sigma &= \eta - \Tilde{E}_0t. 
\end{align}
Although determination of the minimum and maximum eigenvalues is a non-trivial task, we discuss strategies for obtaining these bounds, as well as possible alternatives for scaling the Hamiltonian in \cref{sec:extremumeigenvalues}.
To scale the range of $f(\mathcal{H}^\text{scale})$, we will presently discuss strategies for designing the transform which is one of the main contributions of this work.

% ; Figure~\ref{fig:eigenvalue transform} for example shows the case where a spectrum of eigenvalues (red dots) are adjusted, such that corresponding eigenvalue transforms facilitated by the quantum signal processing is achieved (solid lines superposed in the plot, in this case corresponding to scaled exponential decay functions). This  in QET-U via the $\sigma$ rotation angle on the single-ancilla and the Trotter-evolution time $t$ in the circuits of Figures~\ref{fig:qetu_forrev} and \ref{fig:qetu_for}, as follows;

% ;  This Hamiltonian input model does not require the significant ancilla qubit overhead for performing block-encoding, and may take the form of either (a) a controlled forward-time evolution and anti-controlled reverse-time evolution i.e. Figure~\ref{fig:qetu_forrev}, or (b) a single-ancilla-controlled time evolution unitary i.e. Figure~\ref{fig:qetu_for}. 

% \subsection{Overview of Quantum Signal Processing}

\section{Methods}\label{sec:methods}
We now present the original approaches for imaginary-time evolution investigated in this work, and their applications to three important non-unitary quantum simulation tasks.

\subsection{Exact non-unitary evolution with QET-U}
The normalised imaginary-time or Wick-rotated Schr\"{o}dinger Equation is
\begin{equation} \label{eq:wick-rotated}
    \ket{\psi(\tau)} = A(\tau)e^{-\mathcal{H}\tau}\ket{\psi_0}
\end{equation}
where $\tau = it$ and $A(\tau)=1/\sqrt{\bra{\psi_0}e^{-2\mathcal{H}\tau}\ket{\psi_0}}$ is a normalisation constant.
The main insight of this work is that comparison of the above with \cref{eq:qet-u} shows we can perform ITE on a quantum computer within the QET-U framework.
% by compiling a polynomial Hamiltonian transform which approximates the non-unitary exponential function:
% \begin{equation}
%     P(\mathcal{H}) \approx f(\mathcal{H}) = e^{-\mathcal{H}\tau}.
% \end{equation}

\begin{statement}
\label{statement1}
By compiling phase factors approximating the Hamiltonian function 
\begin{equation} \label{eq:ite_unscaled}
    f(\mathcal{H}^\text{scale}) = \frac{\exp[-\tau (\mathcal{H}^\text{scale}-\sigma)/t]}{\exp[-\tau (E_M^\text{scale}-\sigma)/t ]} 
\end{equation}
(where the scale superscripts represent the scaled Hamiltonian as per \cref{eq: scaled Ham} and $E^\text{scale}_M$ is either $\eta$ or $\frac{\pi}{2}-\eta$), QET-U can be used to apply the exact Wick-rotated Schr\"odinger Equation \cref{eq:wick-rotated} on a quantum computer, up to the accuracy of the polynomial approximation $\mathcal{P}(\mathcal{H})\approx f(\mathcal{H})$.
\end{statement}
By undoing the scaling of $\mathcal{H}^\text{scale}$ at the phase factor optimisation level, the transformation $\exp[-\tau (\mathcal{H}^\text{scale}-\sigma)/t]$ guarantees that the proportion of exponential decay experienced between eigenvalues is identical to  that of the transform $e^{-\mathcal{H}\tau}$ -- this is equivalent to first scaling the input domain of a function, then accordingly scaling the function itself.
As division by constants do not affect the ratio of exponential decay between eigenvalues, division by $\exp[-\tau (E^\text{scale}_M-\sigma)/t]$ and indeed the normalisation constant does not affect the non-unitary dynamics.

We can consider this approach a generalisation of the PITE method~\cite{Kosugi2022} which approximates short ITE steps by also implementing cosine functions of the Hamiltonian using LCU of forward and reverse real-time evolution, facilitated by a stochastic single-ancilla measurement. This strategy of sequentially applying cosine of the Hamiltonian has similarly been explored by others~\cite{PhysRevLett.127.040505, Meister2022} for state preparation. Indeed, the authors of QET-U also note the similarity between the circuit to the Hadamard test~\cite{dong2022ground} which also applies a cosine transform of the Hamiltonian.

However, to achieve a long imaginary time sequence with PITE, one has to sequentially apply PITE stochastic measurements, which can have a low cumulative success probability overall. The authors of PITE initially noted this as a `deplorable drawback' because the overall success probability can be vanishingly small even with less than only 10's of PITE steps~\cite{Kosugi2022}, but have proposed amplitude amplification as a meaningful strategy to counteract this at the expense of a deeper circuit~\cite{Nishi2022}.  

In contrast, because the duration of $\tau$ is determined by the QSP phase factors in QET-U, we can approximately project any amount of ITE using a single ancilla measurement at the end of a QSP sequence. Furthermore, because our approach is performing exact ITE up to a polynomial approximation, it is useful beyond just state preparation, and can be straightforwardly applied to non-unitary evolution. In this work, we opt to use the symmetric phase factors approach described in~\cite{Dong2021Efficient, dong2022ground}, which can find phase factors to accurately approximate polynomials of degree $O(10^4)$. Nevertheless, since ITE is a `lossy' non-unitary process, the probability of a successful ancilla measurement (given by $\mathcal{N}$ in \cref{eq:qet-u}) is expected to decay exponentially if the overlap between the initial state and the projected state decreases, or if $\tau$ is large. We can reason by considering the Taylor expansion of the exponential function that accurately representing a long ITE transformation will require a large degree of the polynomial $P(\mathcal{H})$, which increases the length of the QSP sequence and thus circuit depth.

Our strategy to maximise the probability of successful ancilla projection is to normalise the eigenvalue transform by the exponent of the scaled ground state $\eta$ (if we are performing forward imaginary time) or the maximum excited state $\frac{\pi}{2}-\eta$ (if we are performing reverse imaginary time), motivating the use of $\exp[-\tau (E^\text{scale}_M-\sigma)/t]$ in \cref{eq:ite_unscaled}. This places the burden only on accurately scaling $\mathcal{H}^\text{scale}$ such that \cref{eq:Escale_0} and \cref{eq:Escale_N} hold -- we focus on this challenge in  \cref{sec:extremumeigenvalues}. Assuming this can be achieved, we can then prevent the ground/maximum energy eigenstate from exponential attenuation, and mirrors the ground state preparation strategy proposed in Ref.~\cite{dong2022ground} which we will describe in the next section.

We will also investigate dividing the total imaginary time into $N$ evenly spaced imaginary-time fragments $e^{-\mathcal{H}\tau}=(e^{-\mathcal{H}\Delta\tau})^N$~\cite{Silva2021, dong2022ground}, and each ITE fragment $e^{-\mathcal{H}\Delta\tau}$ is realised stochastically using QET-U. This is of course very similar to PITE and QITE, but the $\Delta\tau$ that can be performed with QET-U is much larger than time-steps that can be performed with PITE and QITE.

\subsection{Ground state preparation}
The most obvious application of ITE is ground state preparation.
Given an initial state $\psi_0 = \sum_i a_i \Phi_i$ (linear combination of Hamiltonian eigenstates $H\Phi_i=E_i \Phi_i$) with nonzero overlap with the ground state $\Phi_0$, the ITE propagator exponentially attenuates the contribution of high energy eigenstates faster than the ground state:
\begin{equation}
    e^{-\mathcal{H}\tau}\ket{\psi_0} = \sum_i a_i e^{-E_i\tau}\ket{\Phi_i},
\end{equation}
thus in the $\tau\rightarrow \infty$ limit the operator prepares the ground state of $\mathcal{H}$. 

The authors of QET-U build on their previous designs of a near-optimal ground state preparation algorithm using LCU and QSP~\cite{Lin2020nearoptimalground} to perform ground state preparation using an eigenstate filtering procedure and QET-U~\cite{dong2022ground}. Given an additional lower bound for the first excited state, they evaluate phase factors which approximately transform the Hamiltonian eigenvalues by an appropriately-scaled Heaviside step function, with the step placed in the spectral gap $\Delta$ between the (scaled) ground and first excited state energies $E^\text{scale}_0$ and $E^\text{scale}_1$:
\begin{equation} \label{eq:heaviside}
    f(\mathcal{H}^\text{scale}) = 
    \begin{cases}
        1 & \text{if } E\leq\mu \\
        0 & \text{if } E>\mu
    \end{cases},
\end{equation}
where $\mu=(E^\text{scale}_0+E^\text{scale}_1)/2$. The authors propose a near-optimal QET-U ground state preparation approach with query complexity $\Tilde{O}(\Delta^{-1}\gamma^{-1}) $ and another with query complexity $\Tilde{O}(\Delta^{-1}\gamma^{-2}) $~\cite{dong2022ground}, where $\gamma$ is the overlap between the initial state and the ground state (the notation $\Tilde{O}(n)$ is $O(n\log^k n)$).
The main drawback of the eigenstate filter is, of course, the need for a bound of the first excited state energy; even an estimate of such an eigenvalue can be challenging to obtain. 
% In practice, although this precomputation might be justified in scenarios where preparing the ground state of a molecule at a single equilibrium geometry is used for subsequent quantum dynamics investigations on a quantum computer (see e.g.~\cite{Chan2022}), it is not ideal for quantum computing electronic structure.

In contrast, performing ITE using QET-U does not require knowledge of the first excited state; it only needs bounds of $E_0$ and $E_{N-1}$ for the QET-U eigenvalue scaling. The complexity of using it for ground state preparation will however still depend on $\gamma$, as a good initial overlap reduces the duration of imaginary time propagation needed for convergence. Furthermore, it can be applied to other tasks which we will discuss presently.

\subsection{Gibbs state and partition function}
% We go beyond ground state preparation of pure quantum states, and investigate the use of QET-U in the non-unitary preparation and evolution of density operators on quantum computers.

% Denisty operators are generalised matrix descriptions of quantum systems that cannot be described by state vectors alone. Given a quantum state which has probability $p_i$ of being in one of the possible normalised (but not necessarily orthogonal) pure states in the set $\{\ket{\psi_i}\}$, the density matrix $\rho$ of such a mixed state is $\rho = \sum_i p_i \ket{\psi_i}\bra{\psi_i}$, and the average expectation value of an observable with matrix operator $\hat{\Omega}$ is $\langle\omega\rangle = \Tr\hat{\Omega}\rho$.
% % If state vectors $\ket{\psi_i}$ can be encoded using $n$-qubits on a quantum computer, the $2^n\times 2^n$ density operator $\rho$ can be also be encoded by stacking its columns (vectorising) and treating it like a state vector $\ket{\rho}$. This in general requires at least two $n$-qubit registers, and does not scale as well as storing wavefunction state vectors. Nevertheless, it can be extremely powerful for computing products of large matrices.
% We will explore two scenarios where this quantum state representation is useful; the preparation of thermal states, and the simulation of open quantum systems. 

The study of finite-temperature systems on quantum computers, in particular computing partition functions and preparing Gibbs states, is considered important tasks relevant to simulation, quantum Boltzmann machine learning and quantum optimisation~\cite{Kosugi2022, Matsumoto_2022}. Different techniques for performing these tasks have been proposed~\cite{Yuan2019theoryofvariational, Wang_2021, Kosugi2022, Matsumoto_2022, Coopmans2022, powers2023exploring}. In this work, we demonstrate that performing ITE with QET-U provides a more natural framework for evaluating thermal properties of quantum states on quantum computers compared with variational~\cite{Yuan2019theoryofvariational, Wang_2021, Matsumoto_2022} and PITE~\cite{Kosugi2022} methods.

The target is to generate the Gibbs state density operator of a system $\rho_\text{sys}(\beta)$  with Hamiltonian $\mathcal{H}_\text{sys}$ at the inverse temperature $\beta=1/k_BT$: 
\begin{equation}
    \rho_\text{sys}(\beta) = \frac{e^{-\mathcal{H}_\text{sys}\beta}}{\Tr e^{-\mathcal{H}_\text{sys}\beta}} = \frac{e^{-\mathcal{H}_\text{sys}\beta}}{Z(\beta)},
\end{equation}
where $k_B$ is the Boltzmann constant and $Z(\beta)$ is the partition function. This can be prepared on a quantum computer by considering a density matrix $\rho$ which is a composite of the system of interest and an `environment' subsystem. We note that this does not model the physical surroundings within which the system is embedded. But as we will show momentarily, it allows us to take the partial trace over the `environment' subsystem $\Tr_\text{env}$ and obtain the reduced density matrix $\rho_\text{sys} = \Tr_\text{env} \rho$.

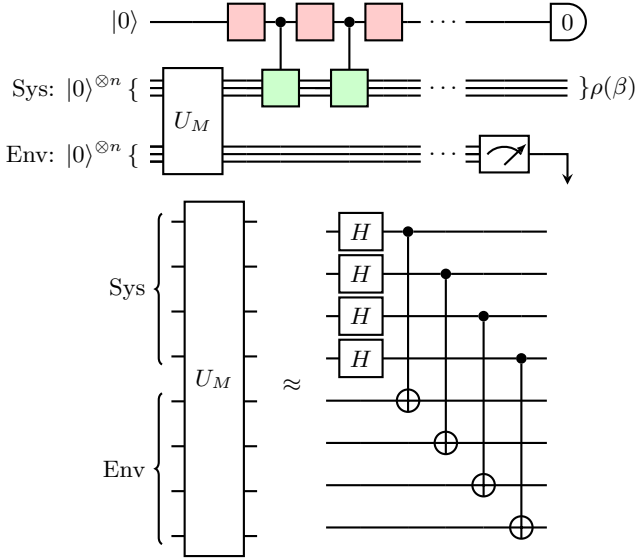
\begin{figure}[h!]
    \centering
    \begin{tikzpicture}
    \node{ 
    \begin{quantikz}[row sep={25pt,between origins}, column sep=5pt]
    &\lstick{$\ket{0}$} & \qw &\gate[style={fill=red!20}]{} & \ctrl{1} & \gate[style={fill=red!20}]{}& \ctrl{1} & \gate[style={fill=red!20}]{}& \qw&\ \ldots\ \qw &\qw &\meterD{0}\\
    &\lstick[wires=1]{Sys: $\ket{0}^{\otimes n} \{$} &\gate[2, bundle={1,2}][0.7cm]{U_M} & \qwbundle[alternate]{}& \gate[style={fill=green!20}]{}\qwbundle[alternate]{}& \qwbundle[alternate]{} &\gate[style={fill=green!20}]{}\qwbundle[alternate]{}& \qwbundle[alternate]{} & \qwbundle[alternate]{} & \ \ldots\ \qwbundle[alternate]{} &\qwbundle[alternate]{} & \qwbundle[alternate]{}\rstick{$\}\rho(\beta)$} & \\
    &\lstick{Env: $\ket{0}^{\otimes n} \{$} & \qwbundle[alternate]{}& \qwbundle[alternate]{}& \qwbundle[alternate]{}& \qwbundle[alternate]{}& \qwbundle[alternate]{}& \qwbundle[alternate]{}&
    \qwbundle[alternate]{}& \ \ldots\ \qwbundle[alternate]{} & \meter{}\qwbundle[alternate]{} & \trash{}
    \end{quantikz}
    };
    \node at (0,-4){
    \begin{quantikz}[row sep={17pt,between origins}, column sep=5pt]
    \lstick[wires=4]{Sys}& \gate[8]{U_M} & \qw\\
    & & \qw\\
    & & \qw\\
    & & \qw\\
    \lstick[wires=4]{Env}& & \qw\\
    & & \qw\\
    & & \qw\\
    & & \qw\\
    \end{quantikz}%
    $\quad\approx$
    \begin{quantikz}[row sep={16pt,between origins}, column sep=5pt]
    & \gate[]{H} & \ctrl{4} &\qw &\qw &\qw &\qw \\
    & \gate[]{H} & \qw& \ctrl{4} &\qw&\qw&\qw \\
    & \gate[]{H} & \qw & \qw & \ctrl{4}&\qw&\qw \\
    & \gate[]{H} & \qw & \qw & \qw&\ctrl{4}&\qw \\
    & \qw & \targ{} &\qw & \qw & \qw&\qw \\
    & \qw & \qw &\targ{} &\qw &\qw&\qw \\
    & \qw & \qw & \qw & \targ{} &\qw&\qw \\
    & \qw & \qw & \qw & \qw &\targ{}&\qw \\
    \end{quantikz}
    };
    \end{tikzpicture}
    
    \caption{Quantum circuit for Gibbs state preparation. The initial state maximally entangles the Sys and Env registers, and QET-U is performed on the Sys register to manifest imaginary-time evolution. Tracing out the Env register yields the Gibbs state density matrix.}
    \label{fig:Gibbs circ}
\end{figure}

We start by generating the maximally entangled state of a bipartite $2n$-qubit register~\cite{Kosugi2022, Matsumoto_2022} using the circuit $U_M$ in the bottom panel of Figure~\ref{fig:Gibbs circ}:
\begin{equation} \label{eq:max entangle}
    \ket{\psi_0} = \frac{1}{\sqrt{2^n}}\sum^{2^n-1}_{i=0}\ket{i}_\text{sys}\ket{i}_\text{env},
\end{equation}
where the first $n$-qubits encode the system of interest with Hamiltonian $\mathcal{H}_\text{sys}$, and the remainder encode the surrounding environment with Hamiltonian $\hat{I}_\text{env}$, making the total Hamiltonian $\mathcal{H}_\text{tot} = \mathcal{H}_\text{sys}\otimes \hat{I}_\text{env}$. 
We wish to evolve this maximally entangled state under the total Hamiltonian $\mathcal{H}_\text{tot}$ by $\tau=\beta/2$:
\begin{align} \label{eq: total ITE}
    \ket{\psi} &= e^{-\mathcal{H}_\text{tot}\tau}\ket{\psi_0} \nonumber\\
    &= \frac{1}{\sqrt{2^n}}\sum^{2^n-1}_{i=0}e^{-(\mathcal{H}_\text{sys}\otimes\hat{I}_\text{env})\tau}\ket{i}_\text{sys}\ket{i}_\text{env}.    
\end{align}
Using a Taylor expansion of the exponential operator, we can show that $e^{-(\mathcal{H}_\text{sys}\otimes\hat{I}_\text{env})\tau}=e^{-\mathcal{H}_\text{sys}\tau}\otimes I_\text{env}$. Therefore, it is only necessary to imaginary time evolve the subsystem $\mathcal{H}_\text{sys}$ which we approximate using QET-U in this work (see Figure~\ref{fig:Gibbs circ}): 
\begin{align} \label{eq:ITE on sys}
    \ket{\psi} =& \frac{1}{\sqrt{2^n}}\sum^{2^n-1}_{i=0}\frac{1}{\sqrt{\mathcal{N}}}P(\mathcal{H}_\text{sys})\ket{i}_\text{sys}\otimes\hat{I}_\text{env}\ket{i}_\text{env} \nonumber\\
    \approx& \frac{1}{\sqrt{2^n}}\sum^{2^n-1}_{i=0}\frac{e^{-\mathcal{H}_\text{sys}\tau}}{\sqrt{\mathcal{N}}}\ket{i}_\text{sys}\otimes\hat{I}_\text{env}\ket{i}_\text{env}
\end{align} 
For now we ignore the QET-U ancilla and assume its measurement yields a successful outcome. The resulting state vector of the $2n$-qubit pure state is approximately the density operator:
\begin{align}
    \rho =& \ket{\psi}\bra{\psi} \nonumber\\
    =& \frac{1}{\mathcal{N}2^n} \sum^{2^n-1}_{i,j=0} (e^{-\mathcal{H}\tau}\ket{i}\bra{j}e^{-\mathcal{H}\tau})_\text{sys} \otimes (\hat{I}\ket{i}\bra{j}\hat{I})_\text{env} \label{eq:gibbpurestate}
\end{align}
(where $e^{-\mathcal{H}_\text{sys}\tau}$ and $\hat{I}_\text{env}$ are both Hermitian). At this point, we can discard the $n$-qubits of the environment register (equivalent to the partial trace over the environment register):
\begin{align}
    \Tr_\text{env}{\rho} =& \frac{1}{\mathcal{N}2^n} \sum^{2^n-1}_{i,j=0} (e^{-\mathcal{H}\tau}\ket{i}\bra{j}e^{-\mathcal{H}\tau})_\text{sys} \otimes \Tr(\ket{i}\bra{j})_\text{env} \nonumber\\
    =& \frac{1}{\mathcal{N}2^n} \sum^{2^n-1}_{i,j=0} (e^{-\mathcal{H}\tau}\ket{i}\bra{j}e^{-\mathcal{H}\tau})_\text{sys} \delta_{ji} \nonumber \\
    =& \frac{1}{\mathcal{N}2^n} \sum^{2^n-1}_{i=0} e^{-\mathcal{H}_\text{sys}\beta}\ket{i}\bra{i},
\end{align}
where the final state on the $n$-qubit system register is now a reduced density operator that is inexpressible as a state vector, which is proportional to the Gibbs state $\rho_\text{sys}$ up to the normalisation factor $\mathcal{N}$ from QET-U.

In \cref{sec:results}, we will justify how if one wishes to extract expectation values of the Gibbs state, the environment register need not be necessary.

We will now show that this normalisation factor, hence the probability of measuring a successful outcome for QET-U, is proportional to the partition function. Reiterating Equation~\ref{eq:qet-u}, the state vector normalisation factor from the polynomial approximation of $e^{-\mathcal{H}_\text{tot}\tau}$ is $\mathcal{N(\tau)} \approx \bra{\psi_0}e^{-2\mathcal{H}_\text{tot}\tau}\ket{\psi_0}$. Combining Equations~\ref{eq:max entangle} and~\ref{eq: total ITE}: 
\begin{align}
    \bra{\psi_0}e^{-2\mathcal{H}_\text{tot}\tau}\ket{\psi_0} &= \frac{1}{2^n}\sum^{2^n-1}_{i,j=0}\bra{i} e^{-2\mathcal{H}\tau}\ket{j}_\text{sys} \bra{i}\hat{I}^2\ket{j}_\text{env} \nonumber\\
    &= \frac{1}{2^n}\sum^{2^n-1}_{i=0}\bra{i} e^{-2\mathcal{H}_\text{sys}\tau}\ket{i} \nonumber\\
    &= \frac{1}{2^n}\Tr\left[e^{-\mathcal{H}_\text{sys}\beta}\right]=\frac{Z(\beta)}{2^n}
\end{align}
shows that, similar to PITE~\cite{Kosugi2022}, the probability of a successful ancilla measurement times the factor $2^n$ gives an estimate of the partition function. However, because of this factor, exponential resources are needed to estimate the partition function. Nevertheless, given these properties, one can calculate e.g. the free energy $F=-\beta^{-1}\ln{Z(\beta)}$, which is important in predicting chemical reaction thermal properties.

\subsection{QET-U for open quantum system dynamics}
The final case of QET-U for ITE studied in this work is in the non-unitary dynamics of open quantum systems. While quantum algorithms for simulating open quantum systems have been proposed (see e.g. Refs.~\cite{Yuan2019theoryofvariational, https://doi.org/10.48550/arxiv.2209.04956, PhysRevA.83.062317, suri2022twounitary}), it has been observed recently that by interleaving real and imaginary time evolution, one can approximate the dynamics of open quantum systems on a quantum computer in an efficient manner (see Refs.~\cite{Kamakari2022, Schlimgen2022}). 

The Lindblad equation models the dynamics of a Markovian quantum system with weak entanglement to the surrounding environment:
\begin{equation}
    \frac{d\rho}{dt} = -i[\mathcal{H},\rho]+\sum_k\left(\hat{L}_k\rho \hat{L}_k^\dagger-\frac{1}{2}\{\hat{L}_k^\dagger \hat{L}_k, \rho \} \right) 
\end{equation}
where $\rho$ is a $2^n\times 2^n$ density operator representation of the quantum system, and $\hat{L}_k$ are Lindblad operators representing the channels that couple the system to the environment; dynamics of the environment is not explicitly captured in this model. The main insight of ~\cite{Kamakari2022, Schlimgen2022} is that the equation can be rewritten in Schr\"odinger form:
\begin{equation}
    \frac{d\ket{\rho}}{dt} = \hat{\mathcal{L}}\ket{\rho},
\end{equation}
with a non-Hermitian Linbladian Hamiltonian:
\begin{align} \label{eq:lindblad}
    \hat{\mathcal{L}} &= -i\hat{I}_A\otimes\mathcal{H}_B + i\mathcal{H}_A^T\otimes\hat{I}_B +\sum_k\hat{L}_k{}_A^*\otimes\hat{L}_k{}_B^{} \nonumber\\
    &\quad - \frac{1}{2}\hat{I}_A\otimes(\hat{L}_k^\dagger \hat{L}_k)_B - \frac{1}{2}(\hat{L}_k^T \hat{L}_k^*)_A\otimes\hat{I}_B \nonumber
\end{align}
admitting the integral form
\begin{equation}
\ket{\rho(t)} = e^{\hat{\mathcal{L}}t}\ket{\rho(0)}.
\end{equation}
The superscripts ${}^T,{}^*,{}^\dagger$ correspond to the transpose, complex conjugate and conjugate transpose respectively.

The `master Hamiltonian' $\hat{\mathcal{L}}$ is composed of operators which act on two effective subsystems (labelled with subscripts $A$ and $B$), similar to the previous thermal state example where the total Hamiltonian is a tensor product of two operators acting on two subsystem Hilbert spaces. It follows that the $2^n\times 2^n$ density matrix $\rho(t)$ of the open system must be represented as a statevector $\ket{\rho(t)}$ across two equally-sized registers of $n$-qubits, and we can consider the time evolution of $\hat{L}$ as operators acting on either the first or second set of $n$-qubits.

As $\hat{\mathcal{L}}$ is typically non-normal, it cannot be embedded using the time evolution oracle directly for QET-U, because its evolution is not unitary.
Fortunately, $\hat{\mathcal{L}}$ admits the following partition into Hermitian and anti-Hermitian parts:
\begin{equation}
    \hat{\mathcal{L}} = i\mathcal{H}_1 + \mathcal{H}_2
\end{equation}
defined by:
\begin{align} 
    \mathcal{H}_1 =& \frac{1}{2i}(\hat{\mathcal{L}}-\hat{\mathcal{L}}^\dagger) \label{eq:H1}\\
    \mathcal{H}_2 =& \frac{1}{2}(\hat{\mathcal{L}}+\hat{\mathcal{L}}^\dagger) \label{eq:H2}.
\end{align}
In the examples considered here, we were able to straightforwardly write down $\mathcal{H}_1$ and $\mathcal{H}_2$ by simply inspecting which operators have an imaginary coefficient, a strategy expected to be scalable.

The time evolution of the `column-stacked' density operator under $\hat{\mathcal{L}}$ can thus be modelled by splitting the unitary and non-unitary evolution and time-stepping:
\begin{align}
     \ket{\rho(t)}&= e^{i\mathcal{H}_1t + \mathcal{H}_2t}\ket{\rho(0)} \\
    &= \left(e^{i\mathcal{H}_1\delta t/2}e^{\mathcal{H}_2\delta t}e^{i\mathcal{H}_1\delta t/2}\right)^N\ket{\rho(0)} + \mathcal{O}(\delta t^2) \label{eq:OQS}.
\end{align}
In this work, the non-unitary time-step is performed using QET-U.
We use qubit measurements to sample the square of the statevector, and reconstruct the density matrix. Because this method of open quantum system dynamics simulation only preserves $\Tr(\rho^2)$, the statevectors are actually divided by $\Tr(\rho)$, trivially obtained once an estimate of the statevector is known.
% Measuring the state of the qubits after time evolution $t$ yields the outer product $\ket{\rho(t)}\bra{\rho(t)}$. 

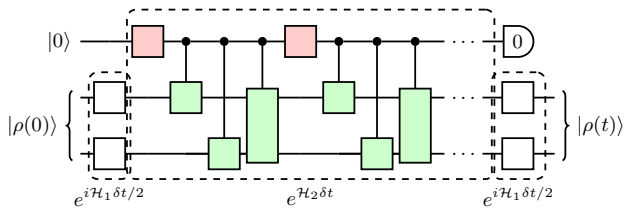
\begin{figure}[h!]
    \centering
    \begin{tikzpicture}
    \node[scale=0.85]{
    \begin{quantikz}[row sep={25pt,between origins}, column sep=9pt]
        \lstick{$\ket{0}$} & \qw &\gate[style={fill=red!20}]{}\gategroup[3,steps=9,style={dashed, rounded corners,inner xsep=2pt}, background,label style={label position=below,anchor=north,yshift=-0.2cm}]{{$e^{\mathcal{H}_2\delta t}$}} & \ctrl{1}& \ctrl{2}& \ctrl{1} & \gate[style={fill=red!20}]{} & \ctrl{1}& \ctrl{2}& \ctrl{1} &\ \ldots\ \qw &\meterD{0}\\
        \lstick[wires=2]{$\ket{\rho(0)}$}& \gate{}\gategroup[2,steps=1,style={dashed, rounded corners,inner xsep=2pt}, background,label style={label position=below,anchor=north,yshift=-0.2cm}]{{$e^{i\mathcal{H}_1\delta t/2}$}} & \qw &\gate[style={fill=green!20}]{}&\qw&\gate[2, style={fill=green!20}]{} & \qw &\gate[style={fill=green!20}]{}&\qw&\gate[2, style={fill=green!20}]{}&\ \ldots\ \qw &\gate{}\gategroup[2,steps=1,style={dashed, rounded corners,inner xsep=2pt}, background,label style={label position=below,anchor=north,yshift=-0.2cm}]{{$e^{i\mathcal{H}_1\delta t/2}$}}&\qw\rstick[wires=2]{$\ket{\rho(t)}$}\\
        & \gate{} & \qw & \qw& \gate[style={fill=green!20}]{}\qw& \qw & \qw & \qw& \gate[style={fill=green!20}]{}\qw & \qw &\ \ldots\ \qw &\gate{}&\qw
    \end{quantikz}
    };
    \end{tikzpicture}
    \caption{Circuit performing dynamics simulation of a vectorised density matrix, representing an open quantum system. QET-U is used to perform the non-unitary time evolution step.}
    \label{fig:lindblad circ}
\end{figure}

\section{Results and Discussion} \label{sec:results}
We perform ITE with QET-U on classically-emulated noise-free quantum circuits. Qubit simulations are developed using the quantum computational chemistry package InQuanto~\cite{inquanto} and pyTket~\cite{Sivarajah_2021}. Benchmark calculations for Hamiltonian diagonalisation and density matrix computations are performed using the SciPy~\cite{2020SciPy-NMeth} package. The Hamiltonian eigenvalues are used as inputs for scaling and shifting the Hamiltonian eigenvalues in QET-U. We highlight this is only for demonstration purposes as in practice the eigenvalues will not be accessible \textit{a priori}. We use the QSPPACK\cite{Dong2021Efficient} package to evaluate the coefficients to the basis of Chebyshev functions, and thus QSP phase factors for approximating a given target polynomial.

\subsection{Ground state preparation}
% \begin{figure}[h!]
%     \centering
%     \begin{tikzpicture}
%         \begin{axis}[domain=0:25,samples=60, ymin=-1, ymax=3, xmin=0.5, xmax=24.5, width=6cm, height=4cm, ticks=none]
%         \addplot[mark=none]{cos(deg(x))}; 
%         \node[label={},circle,draw=black,inner sep=2pt] at (axis cs:3.1,-0.3) {};
%         \node[label={},circle,draw=black,inner sep=2pt] at (axis cs:14.5,0.5) {};
%         \node[label={},circle,draw=black,inner sep=2pt] at (axis cs:22,-0.3) {};
%         \node[label={},circle,draw=black,inner sep=2pt] at (axis cs:17,0.5) {};
%         \draw [-stealth](axis cs:3.1,-0.8) -- (axis cs:3.1,0.2);
%         \draw [-stealth](axis cs:14.5,1) -- (axis cs:14.5,0);
%         \draw [-stealth](axis cs:22,-0.8) -- (axis cs:22,0.2);
%         \draw [-stealth](axis cs:17,0) -- (axis cs:17,1);
%         \node at (axis cs:16,2) {$U$};
%         \node at (axis cs:6,2) {$t$};
%         \end{axis}
%     \end{tikzpicture}
%     \caption{4-site linear Hubbard model}
%     \label{fig:my_label}
% \end{figure}
We study the 4-site, half-filled, linear Fermi-Hubbard chain, which is an 8-qubit problem with the Hamiltonian:
\begin{equation}
    \mathcal{H}=-t\sum_{i,\sigma} \left(\hat{c}^\dagger_{i,\sigma}\hat{c}_{i+1,\sigma} + \hat{c}^\dagger_{i+1,\sigma}\hat{c}_{i,\sigma} \right) + U\sum_i\hat{n}_{i\uparrow}\hat{n}_{i\downarrow},
\end{equation}
where we choose different ratios of on-site electron repulsion $U$ to one-electron hopping term between adjacent sites $t$, representative of weak electron correlation ($U<<t$) and strong electron correlation ($U>>t$) regimes.

\subsubsection{Increasing imaginary time} \label{sec:increaseITE}
We first compile QSP phase factors for 350-, 250-, and 150-degree polynomial approximations of \cref{eq:ite_unscaled}, for $\tau$ between 0 and 5, choosing the Hubbard Hamiltonian where $U=t$. The corresponding QET-U circuits are prepared, where every call to the time evolution oracle uses 25 Trotter steps. We stress that for different values of $\tau$, the only difference is the phase factors used. Each circuit is used to prepare the ground state from a mean-field Hartree-Fock initial state, and the expected energies of each prepared post-selected state are computed.

\begin{figure}[h!]
    \centering
    \includegraphics[width=\linewidth]{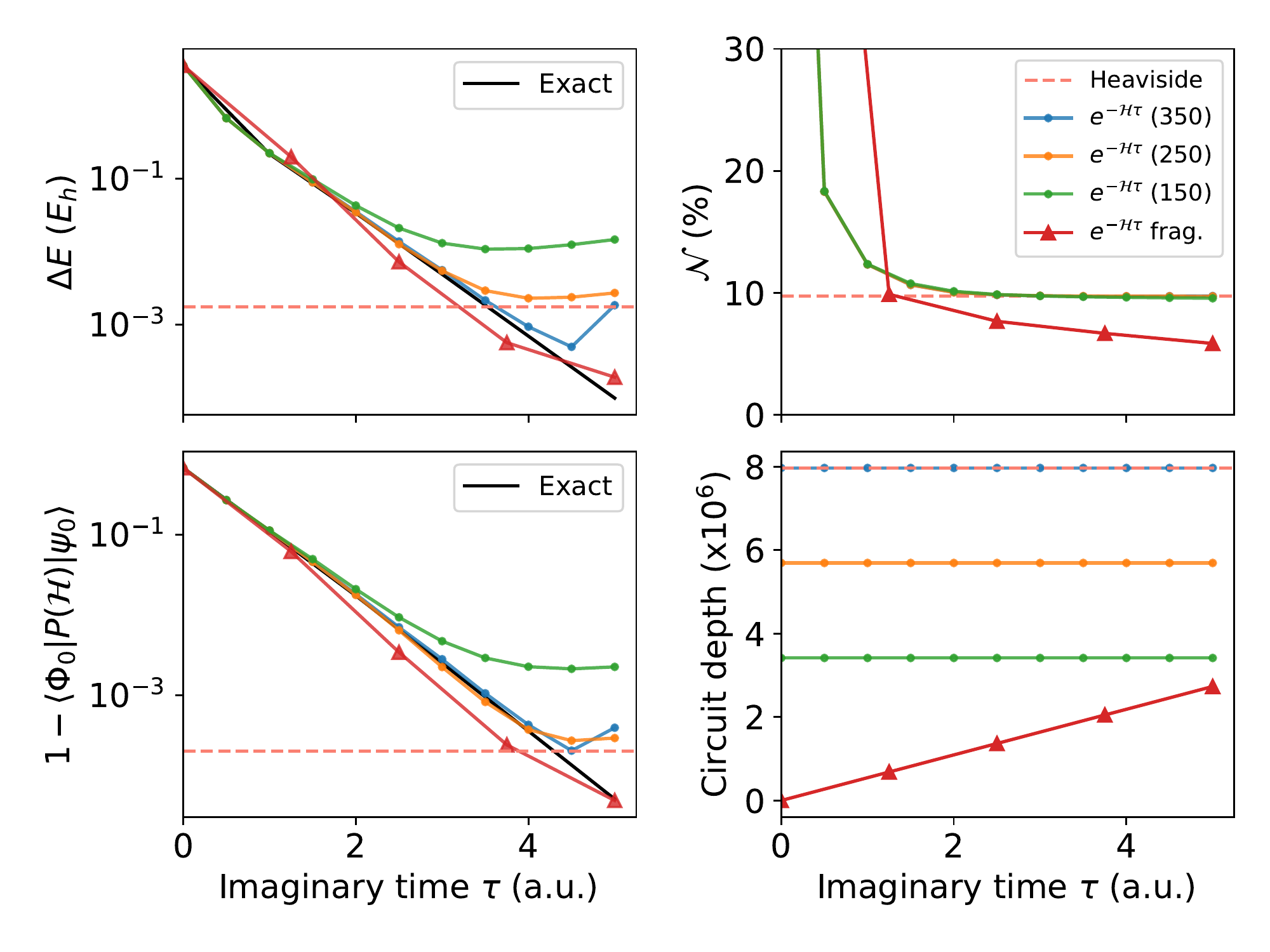}
    \caption{Ground state convergence properties by performing imaginary-time evolution using QET-U. Note that for the fragmented version, the probability of success is the cumulative success probability up to that point in imaginary-time.}
    \label{fig:HubbardITE}
\end{figure}

We find that, in general, the convergence to the ground state is the same as if one is performing exact ITE (see top left panel of \cref{fig:HubbardITE}). However, as expected, for low-degree polynomial approximations of \cref{eq:ite_unscaled}, the energy does not properly converge. As explained previously, this is because at long imaginary times, the sharper exponential decay requires higher order terms to represent accurately. The probability of successfully preparing the post-selected state is displayed in the top right panel of \cref{fig:HubbardITE}, which shows that it levels off around 10\% at long imaginary times.

\subsubsection{Comparison with Eigenstate Filter}
Using the same QET-U circuit, but substituting instead the phase factors for the Heaviside function (\cref{eq:heaviside}) as per Ref.~\cite{dong2022ground}, we were able to prepare a state with an energy that is 1.77m$E_h$ from the exact ground state energy. It similarly shared a 10\% probability of successful post-selection. These numerical results suggest that, for given a circuit depth budget restriction, performing ITE with QET-U can actually provide a better ground state energy convergence than performing eigenstate filtering. 

The strength of the eigenstate filter approach is that it should simply converge to the ground state by increasing the polynomial degree, provided a good first excited state energy is known; in contrast, our ITE approach requires the user to make an informed choice on what is a sufficiently long imaginary time. As we have seen, it may be counterproductive to increase the imaginary time without commensurately increasing the polynomial degree. Therefore, whereas the necessary number of shots for ground state preparation using the eigenstate filter will be determined solely by the success probability, an additional shots overhead might be necessary for probing what is the appropriate duration of imaginary time.

\subsubsection{Imaginary Time `Fragmentation'}
We highlight that within our choice of algorithm parameters, the QET-U circuits required to converge the ground state for this relatively trivial system are considerable.
In an effort to reduce the circuit depth, we explore a fragmentation strategy proposed in Ref.~\cite{Silva2021}. From a QSP perspective, we can also choose lower degree polynomial approximations compared with earlier demonstrations to perform each imaginary-time fragment. Unlike the above experiments though, the metric for probability of preparing a target state is now the cumulative probability of success in applying each ITE fragment; in the case of PITE, this total probability has been found to be vanishingly small even after a small number of PITE steps~\cite{Kosugi2022}. The circuit depth also is no longer constant but will grow linearly with the number of fragments. However, we will see momentarily that because the cost of performing each time step can be reduced by choosing a lower degree approximation of the transform, the total cost can actually be less than the previous strategy.

We divide the total imaginary-time into 4 time steps ($\Delta\tau=1.25$), and each imaginary-time step is approximated with only a 30-degree polynomial. The top panels of \cref{fig:HubbardITE} show the energy convergence behaviour and the cumulative success probability of the sequential QET-U ancilla measurements, respectively. The results suggest that fragmentation offers a similar performance in terms of convergence and probability of success. Critically, the final circuit depth of performing these three QET-U imaginary-time fragments is less than half that is necessary to perform the 350-degree, single-shot imaginary-time QET-U from \cref{sec:increaseITE}. This strategy of performing fragmentation hybridises the best features of PITE and QET-U, and can be a strong strategy for ground state preparation in terms of success probability, circuit depth, minimum pre-computation cost and energy convergence.

\subsubsection{Severe dependence on the initial trial state}
While we have shown that the method can converge to the ground state, two other important metrics are the circuit depth and the probability of success. Using the same Hamiltonian as before, we perform the transform of \cref{eq:ite_unscaled} exactly using all possible 4-body single determinant initial reference states, and confirm that the probability of success scales as the square of the initial overlap $O(\gamma^2)$ -- see Appendix. While this is expected from \cref{eq:qet-u}, it does highlight the importance of selecting an appropriate initial state -- particularly in the Fermi-Hubbard model, which is popular for describing strongly-correlated systems (in the $U>>t$ regime). In these regimes, the overlap between the single reference mean-field state and the ground state is significantly smaller. As a result, the probability of success can vanish. To show this, we present in the Appendix exact simulations of the fragmented ITE using the transform \cref{eq: scaled Ham} for values of $U/t=$5, 10, 20, 40, which confirm that to reach a given energy accuracy the probability of success can become prohibitively small. Our results echo previous conclusions on the severe dependence of the initial state for quantum state preparation, and further highlight the need for improved initial state preparation strategies suitable for strongly correlated systems. It is however possible to boost the probability of success using amplitude amplification, at the cost of deepening the circuit. How long it takes for a state to converge, and thus the required circuit complexity, appears to depend less loosely on the initial trial state; we find that states with unremarkable initial overlaps can converge to a given level of accuracy much quicker than one with a high overlap. However, the probability of success will be the limiting factor.

\subsection{Gibbs state preparation}
For Gibbs state preparation and computation of the partition function, we test the 4-site transverse-field Ising model with the Hamiltonian:
\begin{equation}
    \mathcal{H}=-J\sum_i Z_iZ_{i+1} +g\sum_iX_i,
\end{equation}
where we also choose different ratios of the nearest neighbour spin interaction $J$ and the coupling strength of the external transverse field $g$, exploring the $g<<1$ to $g>>1$ regimes. We choose inverse temperature $\beta$ to be 0.5, 1.0 and 2.0. 8-qubits and 50-degree polynomial approximations are used to prepare the corresponding Gibbs states, and each call to the controlled time evolution unitary is divided into 8 steps. 
We measure the expected magnetisation $M=\sum_iZ_i$ of the prepared density matrix, as well as the success probability.

\begin{figure}[h!]
    \begin{centering}
    \includegraphics[width=.8\linewidth]{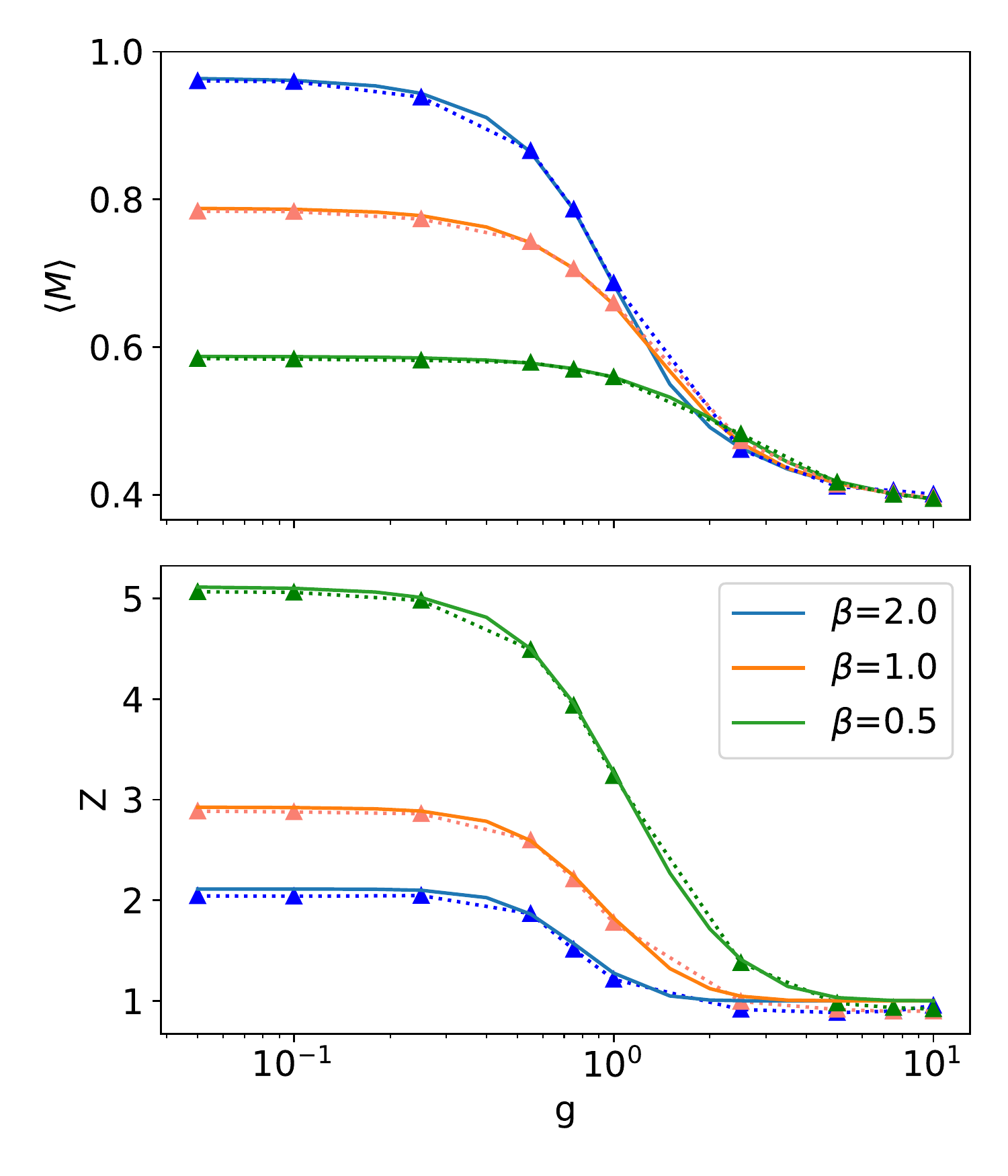}
    \end{centering}
    
    \caption{Gibbs state preparation and partition function calculation using QET-U. Solid lines represent exact values, and markers with dotted lines show the values generated using QET-U emulation. The top panel shows the magnetization computed from the prepared Gibbs state. The bottom panel shows the success probability multiplied by $2^4$.}
    \label{fig:Gibbs}
\end{figure}

The density matrices of thermal states prepared by QET-U have magnetisation values and partition functions that agree well with exact values across the regimes tested. For such a small system, the circuit depth required to converge across the range of $g$ studied is 6750 gates; the $2n$ qubit cost and the exponential cost in partition function estimation means that more work is needed to reap any potential quantum advantage in thermal quantum simulation.

We highlight that it is possible to reduce the qubit cost to $n$ if we extract only thermal shadows of a Gibbs state~\cite{Coopmans2022}, where QET-U would be equally applicable. Furthermore, we note from \cref{eq:gibbpurestate} that if we are performing standard sampling of the mixed Gibbs state or measuring the expectation value of a unitary (as we are doing so here), the second register is actually not necessary;  this is because if we measure the second `Env' register in the binary state $\ket{b_k}$, the `Sys' register will also be in state $\ket{b_k}$ -- which can also be obtained by randomly sampling $\ket{b_k}$ states from a $n$-qubit register. The second register is only necessary for algorithms that require the use of a purified state, such as coherent amplitude estimation.

\subsection{Open quantum system dynamics}
We consider the basic two-level open quantum system with the Hamiltonian:
\begin{equation}
    \mathcal{H}= -\frac{\delta}{2}\hat{\sigma}_z -\frac{\Omega}{2}\hat{\sigma}_x
\end{equation}
and a single amplitude damping channel:
\begin{equation}\label{eq:ampdamp}
    \hat{L}=\sqrt{\gamma}\hat{\sigma}_-=\sqrt{\gamma}(\hat{\sigma}_x-i\hat{\sigma}_y),
\end{equation}
where $\delta$ is the detuning parameter, $\Omega$ is the Rabi frequency and $\gamma$ is the damping constant or spontaneous emission rate.
The density matrix of this system is a $2\times2$ matrix, therefore its vectorised form can be encoded on a $2^2$-element state vector, corresponding to 2 qubits which we will respectively label $A$ and $B$. 

If we first consider the case where $\delta=\Omega=0$, then
\begin{equation}
    \hat{\mathcal{L}} = \hat{L}_A^*\otimes\hat{L}_Bt - \frac{1}{2}\hat{I}_A\otimes(\hat{L}^\dagger\hat{L})_Bt - \frac{1}{2}(\hat{L}^T \hat{L}^*)_A\otimes\hat{I}_Bt.
\nonumber
\end{equation}
Because \cref{eq:ampdamp} is a real matrix, $\hat{L}^*=\hat{L}$ and $\hat{L}^\dagger=\hat{L}^T$, and therefore,
\begin{align}
    \hat{L}_A\otimes\hat{L}_B &= \gamma(\hat{\sigma}_x{}_A\otimes\hat{\sigma}_x{}_B-i\hat{\sigma}_x{}_A\otimes\hat{\sigma}_y{}_B \nonumber\\
    &\quad -\hat{\sigma}_y{}_A\otimes\hat{\sigma}_y{}_B-i\hat{\sigma}_y{}_A\otimes\hat{\sigma}_x{}_B) \nonumber
\end{align}
and
\begin{align}
    \hat{L}^\dagger\hat{L} &= \gamma(\hat{I}-i\hat{\sigma}_x\hat{\sigma}_y{}+i\hat{\sigma}_y\hat{\sigma}_x+\hat{I}) \nonumber\\
    &= 2\gamma(\hat{I}+\hat{\sigma}_z). \nonumber
\end{align}
We can therefore determine the Hermitian and anti-Hermitian components \cref{eq:H1} and \cref{eq:H2} of the master Hamiltonian: 
\begin{align}
    \mathcal{H}_1 &= -\gamma(\sigma_{xA}\otimes\sigma_{yB} + \sigma_{yA}\otimes\sigma_{xB}) \nonumber \\
    \mathcal{H}_2 &= \gamma(\sigma_{xA}\otimes\sigma_{xB} - \sigma_{xA}\otimes\sigma_{xB} - \sigma_{zA} - \sigma_{zB} - 2\hat{I}). \nonumber
\end{align}
In this case, the ITE of $\mathcal{H}_2$ must reflect the exact non-unitary evolution, and the imaginary-time is reversed. We therefore use QET-U to perform \cref{eq:ite_unscaled} which undoes the scaling, with $\tau>0$ and normalised to the maximum eigenvalue. We use only a 6th degree polynomial to approximate the transform, and two Trotter steps for each call of the time evolution oracle in the QSP sequence. We therefore perform time evolution using a circuit implementation of \cref{eq:OQS}, for a total time of 10~a.u. with $\delta t=1.0$~a.u.. 

\begin{figure}[h!]
    \centering
    \includegraphics[scale=0.45]{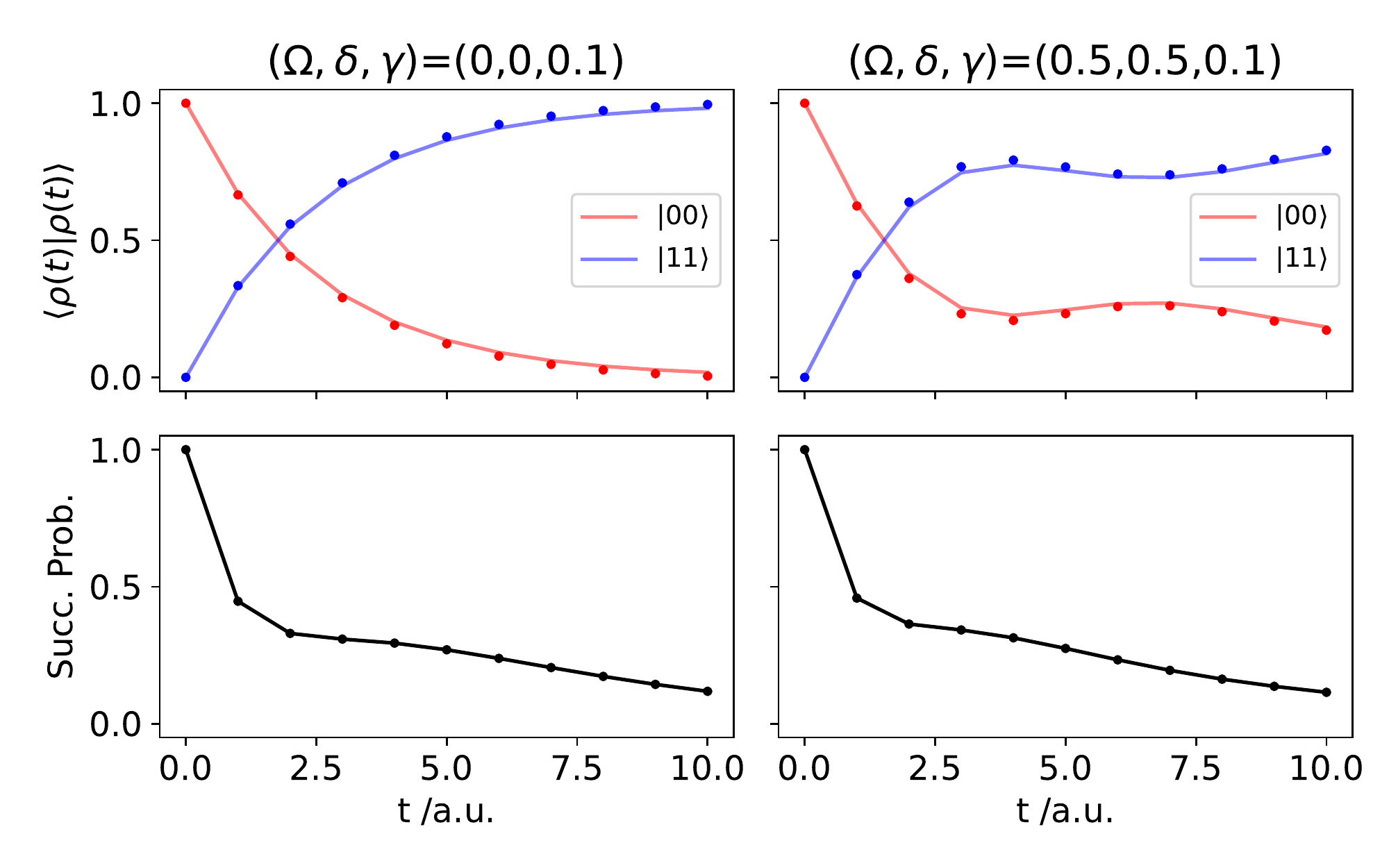}
    \caption{Propagating open quantum system dynamics using QET-U.  The populations of the ground and excited states are encoded as the qubit amplitudes, and in practice will be obtained by sampling. The success probability is cumulative, and for the time scales we have explored remains above 10\%.}
    \label{fig:OQS}
\end{figure}

In this scenario, the populations of the excited state and ground state are given by the square roots of the probabilities of measuring the $\ket{00}$ and the $\ket{11}$ states respectively. These are plotted in \cref{fig:OQS}.  The populations agree well with the exact solution.
Reintroducing the Hamiltonian dynamics, which involves only two additional terms to $\mathcal{H}_1$:
\begin{equation*}
    \mathcal{H}_1 = -\gamma(\sigma_{xA}\otimes\sigma_{yB} + \sigma_{yA}\otimes\sigma_{xB})
\end{equation*}
produces evolution that agrees well with the exact propagation.

Because the non-unitary dynamics has to be Trotterised, more shots will be required for long time evolution as the cumulative success probability decays. In this simple test case as shown in \cref{fig:OQS}, the cumulative success probability remains above 10\% even at long times. At each time step, the circuit depth is 904 gates; predicting the total 10 a.u. dynamics described herein requires a gate depth of $O(10^4)$.

\subsection{Hamiltonian scaling: Bounds of extremum eigenvalues} \label{sec:extremumeigenvalues}

The QET-U approach we used here requires pre-computation of the extremum eigenvalues, which is necessary for the eigenvalue scaling (see \cref{eq: scaled Ham}). For our demonstrations, we compute exactly the values $E_0$ and $E_{N-1}$ for eigenvalue shifting and scaling in our QET-U numerics. In general, this is of course an extremely challenging task and estimating good bounds is dependant on the applicability of this algorithm. We will conclude by considering some strategies to address this step.

The simplest approach for calculating the approximate eigenvalue bounds of the operator when it is supplied as a linear combination of terms $\mathcal{H} = \sum_k a_k \mathcal{P}_k$ (which is the case for second quantised chemistry and physics Hamiltonians) is given by
\begin{equation}
    E_\pm = \pm \sum_k |a_k| \cdot \norm{\mathcal{P}_k}.
\end{equation}
Where the $\norm{\mathcal{P}_k}$ spectral norm of operator $\mathcal{P}_k$. Other methods to find the lower bound of the ground state $\Tilde{E}_0$ can be provided by computing the energy of the initial reference state $\Tilde{E}_0 = \bra{\psi_0}\mathcal{H}\ket{\psi_0}$. The accuracy of this bound naturally depends on the overlap between the initial state and the ground state; this holds its own unique challenges in the strong correlation regime as discussed. An upper bound of the maximum eigenvalue $\Tilde{E}_{N-1}$ can in turn be estimated using Gershgorin's Circle Theorem~\cite{GoluVanl96}, whereby discs bounding eigenvalues on the complex plane are constructed using diagonal elements of $\mathcal{H}$ (the origins of said discs) and sum of off diagonal elements on a corresponding row (the radius of said discs). The fact that eigenvalues of $\mathcal{H}$ are real and symmetric can be exploited to simplify upper bound determination. 
Furthermore, like the LCU Hamiltonian input model, it should be possible to divide the coefficients of Pauli terms by the norm of the Hamiltonian, and perform controlled time evolution of this Hamiltonian instead.

\section{Conclusions} \label{sec:conclusion}

We show that QET-U can be leveraged to perform non-unitary evolution on early fault-tolerant quantum computers using only a single additional qubit. By combining recent advances in efficient computation of phase factors, we were able to generate high-degree polynomial approximations of the ITE propagator.
While normalisation of the ITE propagator gave a boost to the success probability for the systems tested, as we scale to more qubits or systems with strong correlation, the probability of success can become prohibitively small to offer quantum speedup. As mentioned, a natural extension to counteract this is to incorporate amplitude amplification, at the expense of additional quantum gates~\cite{Nishi2022}. Hybridising this approach with other cheap state preparation algorithms or initial states to `warm-start' the ITE would also improve the probability of success. In combination with appropriate scheduling in the fragmentation approach~\cite{Silva2021}, QET-U for ITE may prove powerful in the early fault-tolerant domain. Furthermore, a more concrete understanding of how errors in the Hamiltonian rescaling might influence the error will be valuable.

The utility of our described approach will ultimately depend on the non-unitary simulation task that the user wishes to perform. For example, if one already has access to an accurate estimate of the ground state or the composition of eigenstates in the initial state, well established methods like iterative phase estimation might be more useful than the methods explored here for ground state preparation. Applications to open quantum system dynamics investigated here seem promising for large scale Lindbladian dynamics simulations, though the diminishing cumulative probability of success needs to be resolved.  
QET-U for ITE can trivially be translated to first-quantized e.g. real-space grid quantum simulations for ground state preparation~\cite{Kosugi2022, Chan2022}, and indeed even to molecular geometry optimisation as recently proposed in Ref.~\cite{Kosugi2022Geo}. Extension to general time-integration of non-unitary differential equations is obvious.

\section*{Acknowledgements} 
The authors would like to thank Alexis Ralli for helpful discussions, and Michelle Sze, Marcello Benedetti for feedback on the manuscript. H.H.S.C. would like to thank Richard Meister, Sam McArdle, B\'alint Koczor and Simon Benjamin for helpful discussions. H.H.S.C. is supported by the Croucher Foundation, Hong Kong.

\bibliographystyle{apsrev4-1.bst}
\bibliography{references.bib}

\onecolumngrid
\appendix*
\section{} \label{sec:appendix}

\subsection{Alternative methods for imaginary time evolution on quantum computers}
\noindent
We provide brief overviews of the three classes of quantum algorithms that approximate ITE.

The quantum imaginary time evolution (QITE) algorithm is perhaps the most widely discussed in the literature. It uses Pauli tomography to approximate non-unitary imaginary-time propagation with Trotterized unitary evolution. Many variations and improvements of the QITE have since been proposed. The method is not probabilistic, but requires classical computing resources for tomography.

The variational imaginary time evolution (VITE) is a general class of hybrid quantum-classical approaches, where a parameterised quantum circuit ansatz is optimised classically such that the ansatz convergence mimics the action of ITE. This method is also not probabilistic, but requires classical computing resources for variational optimisation.

Probabilistic imaginary time evolution (PITE) achieves non-unitary evolution by entangling real time evolution of an $n$-qubit system to an ancilla qubit, and triggering a partial collapse of the $n$-qubit state by measuring the ancilla qubit. A successful outcome prepares a function proportional to $\cos(I-H\delta t)$, thereby approximating to first order the short-ITE operator for a small time step. We discuss this method in greater detail in the main text.

\subsection{QSP and QET-U}
% Quantum Signal Processing is a class of methods for preparing matrix functions on a quantum computer. It was recognised that quantum algorithms with known speedups over their classical analogues can be expressed as matrix function operations, and therefore be unified under a QSP formulation or its generalisation the Quantum Singular Value Transform (QSVT).

The action of the QSP sequence can be interpreted as a set of orthogonal rotations in the Bloch sphere of a single qubit~\cite{Chuang2021}. We define a signal operator which is an $R_z$ rotation by a fixed angle $\theta$, and a set of signal processing operators which are $R_x$ rotations each with different phase factors $\{\varphi_i\}$. The resulting amplitude of the $\ket{0}$ state upon application of this QSP sequence $\bra{0}U(\theta)\ket{0}$ is a linear combination of $\cos(2k\theta)$ (top left block of $U(\theta)$). Rewriting this expression using the definition of Chebyshev polynomials of the first kind $\cos(k\theta)=T_k(\cos\theta)$ shows that it is in fact an even polynomial $P(\theta)$ in the basis of Chebyshev polynomials:
\begin{equation}
    P(\theta) = \sum^{d/2}_{k=0}c_kT_{2k}(\cos\theta), 
\end{equation}
where the coefficients $c_k$ are determined by the phase factors, and the degree $d$ is determined by the number of phase factors in the QSP sequence. 

\begin{figure}[h!]
    % \begin{subfigure}[b]{.45\linewidth}
    \includegraphics[width=0.3\linewidth]{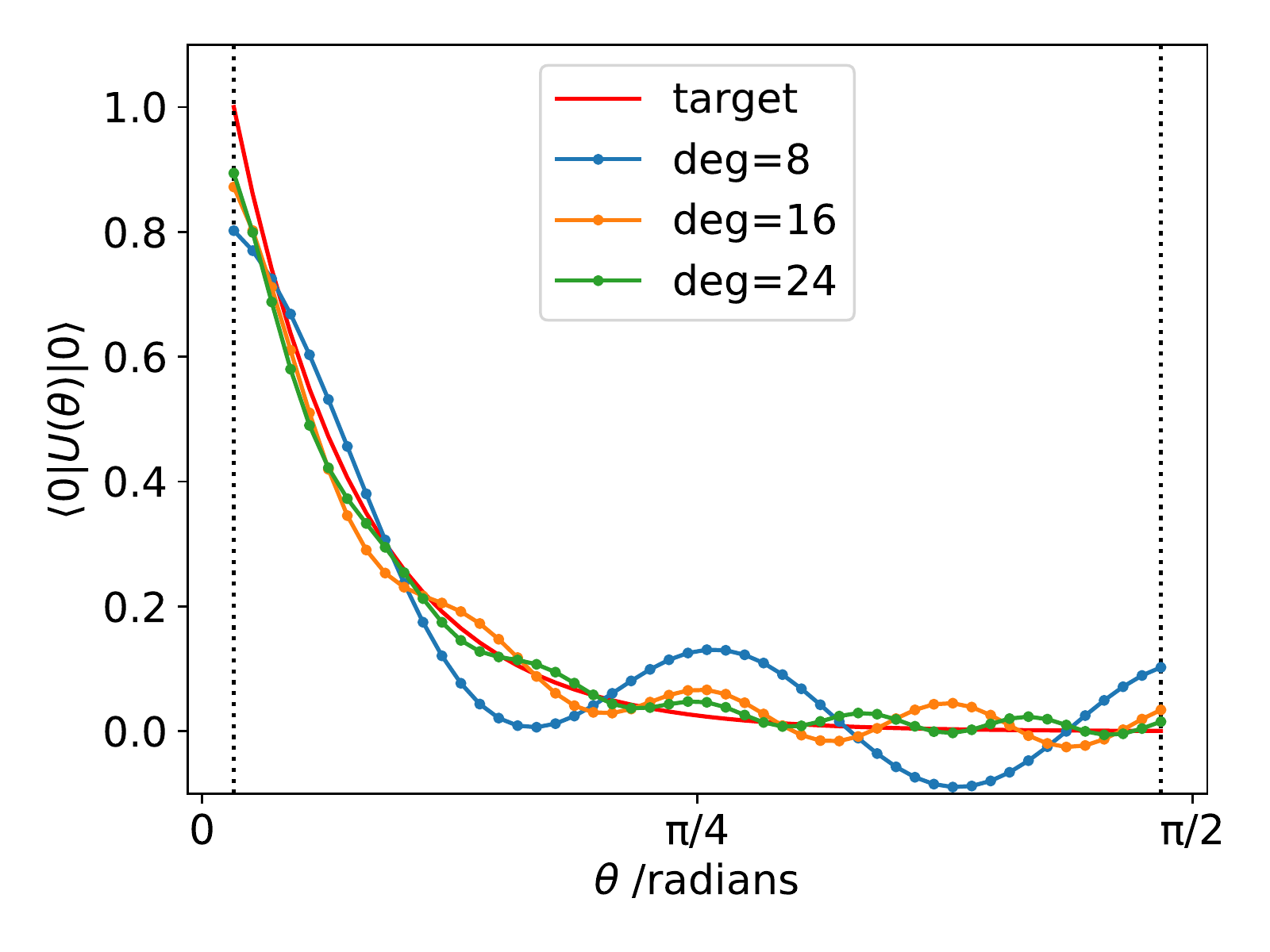}%
    \includegraphics[width=0.3\linewidth]{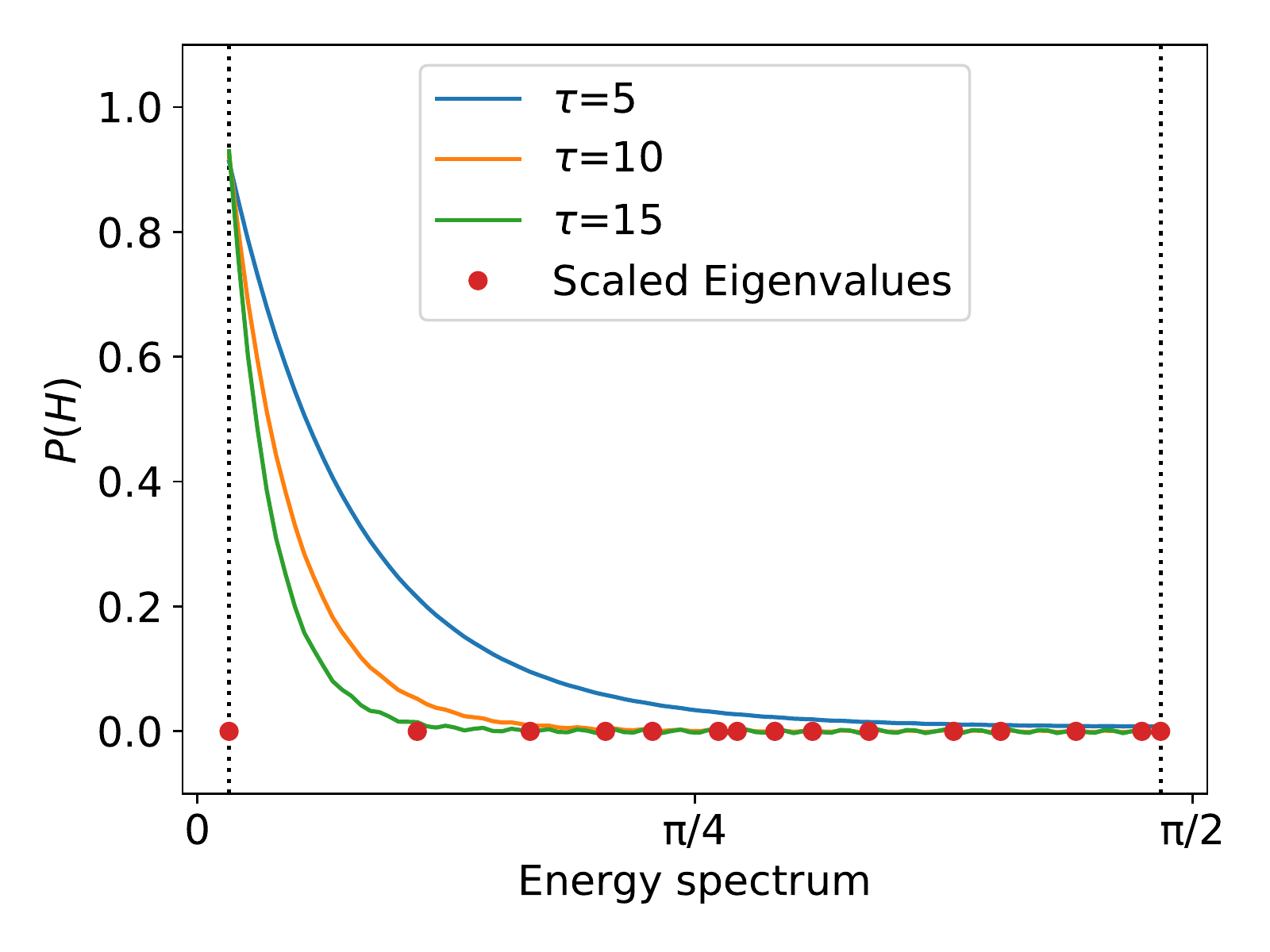}
    \caption{The exponential decay transform of eigenvalues}\label{fig:eigenvalue transform}
\end{figure}

Evaluation of the phases given a target transform can be extremely efficient, especially in the case of a symmetric set $\{\varphi_0, \varphi_1, \varphi_2 \dots \varphi_2, \varphi_1, \varphi_0 \}$~\cite{Dong2021Efficient}, which we use here. 
The left plot in Figure~\ref{fig:eigenvalue transform} shows cases where phase factors are compiled such that $\bra{0}U(\theta)\ket{0}$ approximates a normalised exponential decay target function $e^{-\theta\tau}$, for $\theta \in [\eta, \frac{\pi}{2}-\eta]$, at three different degree polynomial approximations. The small constant term $\eta$ is used to provide numerical stability when evaluating phase factors~\cite{dong2022ground}.

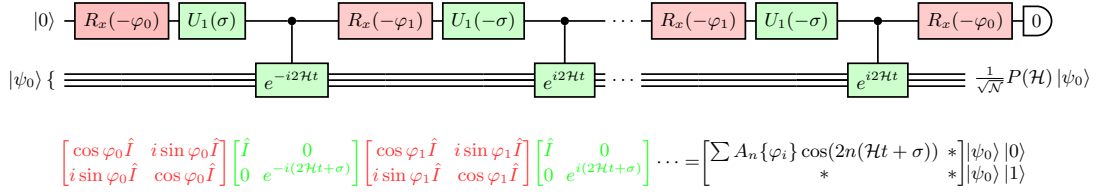
\begin{figure}[h!]
    \centering
    \begin{tikzpicture}
    \node [scale=0.8]{
    \begin{quantikz}[row sep={28pt,between origins}, column sep=5pt]
    &\lstick{$\ket{0}$} & \gate[style={fill=pink}]{R_x(-\varphi_0)} & \gate[style={fill=green!20}]{U_1(\sigma)} & \ctrl{1} & \gate[style={fill=red!20}]{R_x(-\varphi_1)} & \gate[style={fill=green!20}]{U_1(-\sigma)} & \ctrl{1} & \ \ldots\ \qw & \gate[style={fill=red!20}]{R_x(-\varphi_1)} & \gate[style={fill=green!20}]{U_1(-\sigma)} & \ctrl{1} & \gate[style={fill=red!20}]{R_x(-\varphi_0)} &\meterD{0} \\
    &\lstick[wires=1]{$\ket{\psi_0} \{$} &\qwbundle[alternate]{}&\qwbundle[alternate]{}&\gate[style={fill=green!20}]{e^{-i2\mathcal{H}t}}\qwbundle[alternate]{}& \qwbundle[alternate]{}&\qwbundle[alternate]{} &\gate[style={fill=green!20}]{e^{i2\mathcal{H}t}}\qwbundle[alternate]{}  & \ \ldots\ \qwbundle[alternate]{} & \qwbundle[alternate]{} & \qwbundle[alternate]{} &\gate[style={fill=green!20}]{e^{i2\mathcal{H}t}}\qwbundle[alternate]{} &  \qwbundle[alternate]{}\rstick{$\frac{1}{\sqrt{\mathcal{N}}}P(\mathcal{H})\ket{\psi_0}$}
    \end{quantikz}
    };
    \node at (0, -1.5)[scale=0.8]{$
    \textcolor{red!80}{\begin{bmatrix} 
    \cos\varphi_0\hat{I} & i\sin\varphi_0\hat{I} \\
    i\sin\varphi_0\hat{I} & \cos\varphi_0\hat{I} \\
    \end{bmatrix}}
    \textcolor{green!90}{\begin{bmatrix} 
    \hat{I} & 0 \\
    0 & e^{-i(2\mathcal{H} t+\sigma)} \\
    \end{bmatrix}}
    \textcolor{red!80}{\begin{bmatrix} 
    \cos\varphi_1\hat{I} & i\sin\varphi_1\hat{I} \\
    i\sin\varphi_1\hat{I} & \cos\varphi_1\hat{I} \\
    \end{bmatrix}}
    \textcolor{green!90}{\begin{bmatrix} 
    \hat{I} & 0 \\
    0 & e^{i(2\mathcal{H} t+\sigma)} \\
    \end{bmatrix}}\dots
    =
    \begin{blockarray}{ccc}
    \begin{block}{[cc]c}
      \sum A_n\{\varphi_i\}\cos(2n(\mathcal{H} t + \sigma)) & * & \ket{\psi_0}\ket{0} \\
      * & * & \ket{\psi_0}\ket{1} \\
    \end{block}
    \end{blockarray}
    $};
    \end{tikzpicture}
   
    \caption{QET-U for imaginary time evolution}
    \label{fig:qetu_for}
\end{figure}

We highlight that an alternative construction of QET-U, which calls the controlled time evolution operator only once at every instance of the signal operator (see Figure~\ref{fig:qetu_for}), achieves the same polynomial transformation with reduced circuit depth~\cite{Dong2021Efficient}. This is the original version as proposed by the authors of QET-U, which we also use.

In the single-qubit QSP scenario, the set of symmetric phase factors $\{\varphi_i\}$ is optimised such that the desired transform of the $\ket{0}$ amplitude is achieved for $\theta$ between a range of angles $[\eta, \frac{\pi}{2}-\eta]$. It is clear then that to manifest this same transform on the eigenvalues of a Hamiltonian in QET-U, the Hamiltonian must be scaled and shifted such that its energy spectrum fits between those same values for which the desired transformation is achieved. Figure~\ref{fig:eigenvalue transform} also shows the case where a spectrum of eigenvalues (red dots) is adjusted, such that corresponding eigenvalue transforms facilitated by the quantum signal processing are achieved (solid lines superposed in the plot, in this case corresponding to scaled exponential decay functions).

\begin{figure}[h!]
    \centering
    % \end{subfigure}%
    % \vspace{0.5cm}
    % \begin{subfigure}[b]{\linewidth}
    % \caption{Controlled, Trotterized time evolution}\label{fig:ctrl Trot}
    \begin{tikzpicture}
    \node[scale=0.7]{    
    \begin{quantikz}[row sep={10pt,between origins}]
    & \ctrl{1} & \qw \\ [0.3cm]
    & \gate[4]{e^{-i\mathcal{H}t}} & \qw \\
    & & \qw\\
    & & \qw\\
    & & \qw\\
    \end{quantikz}%
    $\quad=$
    \begin{quantikz}[row sep={10pt,between origins}]
    & \ctrl{1} & \ctrl{1} & \push{\quad\dots\quad} &\ctrl{1} & \qw \\ [0.3cm]
    & \gate[4]{e^{-i\mathcal{H}\delta t}} & \gate[4]{e^{-i\mathcal{H}\delta t}} &\push{\quad\dots\quad} & \gate[4]{e^{-i\mathcal{H}\delta t}} & \qw \\
    & & &\push{\quad\dots\quad} & & \qw\\
    & & &\push{\quad\dots\quad} & & \qw\\
    & & &\push{\quad\dots\quad} & & \qw
    \end{quantikz}
    };
    \node at (0,-2) [scale=0.6]{
    \begin{quantikz}[row sep={18pt,between origins}, column sep={10pt}]
    & \ctrl{1} & \qw \\ [0.3cm]
    & \gate[4]{e^{-i\mathcal{H}\delta t}} & \qw \\
    & & \qw\\
    & & \qw\\
    & & \qw\\
    \end{quantikz}%
    $\quad\approx$
    \begin{quantikz}[row sep={18pt,between origins}, column sep={3pt}]
    & \qw &\qw & \qw & \qw  & \ctrl{3}  & \qw & \qw & \qw & \qw & \qw & \qw &\qw & \qw & \qw  & \ctrl{4}  & \qw & \qw & \qw & \qw & \qw &\rstick{\dots}\\ [0.3cm]
    & \qw & \ctrl{1} & \qw & \qw &\qw & \qw & \qw & \ctrl{1} &\qw & \qw & \gate{R_x(\frac{3\pi}{2})} & \ctrl{1} & \qw & \qw &\qw & \qw & \qw & \ctrl{1} &\gate{R_x(\frac{\pi}{2})} & \qw & \rstick{\dots}\\
    & \qw &\targ{}  & \ctrl{1} & \qw & \qw & \qw & \ctrl{1}  & \targ{} & \qw & \qw & \gate{H} &\targ{}  & \ctrl{1} & \qw & \qw & \qw & \ctrl{1}  & \targ{} & \gate{H} & \qw & \rstick{\dots}\\
    & \gate{H} &\qw & \targ{}  & \qw & \gate{R_z(\theta)} & \qw & \targ{} & \qw & \gate{H} & \qw & \gate{H} &\qw & \targ{}  & \ctrl{1} & \qw & \ctrl{1} & \targ{} & \qw & \gate{H} & \qw & \rstick{\dots}\\
    & \qw &\qw & \qw & \qw  & \qw & \qw  & \qw & \qw & \qw & \qw & \qw &\qw & \qw & \targ{}  & \gate{R_z(\theta)} & \targ{}  & \qw & \qw & \qw & \qw & \rstick{\dots}
    \end{quantikz}
    };
    \end{tikzpicture}
    \caption{Controlled, Trotterised time evolution. Because Pauli gadgets are used to implement the exponent of Pauli operators, only control to the $R_z(\theta)$ is necessary.}
    \label{fig:ctrl Trot}
\end{figure}
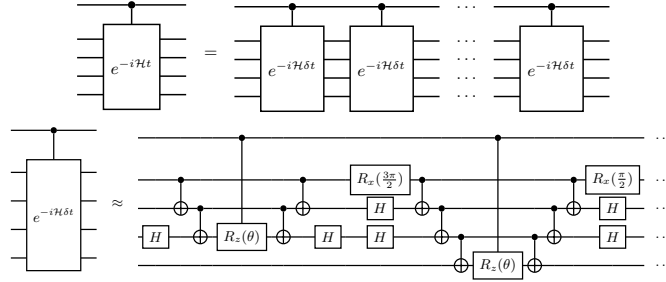

Another consideration is that, in practice, performing forward and reverse time evolution unitaries will require Trotter approximation.  In this work, we use second-quantized Hamiltonians which lend themselves to Pauli gadgets that can be entangled to the ancilla with only one control rotation gate (see~\cref{fig:ctrl Trot}), but highlight that the framework is also applicable to first-quantized Hamiltonians.

% \begin{figure}[h]
% \centering
% \begin{quantikz}
% \lstick{$\ket{0}$} &  \gate{R_y(\theta_0)} &\ctrl{1} \gategroup[4,steps=4,style={dashed,
% rounded corners,fill=blue!20, inner xsep=3.5pt},
% background,label style={label position=below,anchor=
% north,yshift=-0.2cm}]{{ x d}} & \qw & \qw &\gate{R_y(\theta_4)} &\qw  \\
% \lstick{$\ket{0}$} &  \gate{R_y(\theta_1)}  & \targ{}  & \ctrl{1} &\qw &\gate{R_y(\theta_5)} & \qw \\
% \lstick{$\ket{0}$} &  \gate{R_y(\theta_2)}  & \qw & \targ{} & \ctrl{1}  & \gate{R_y(\theta_6)} & \qw \\
% \lstick{$\ket{0}$} &  \gate{R_y(\theta_3)}  & \qw & \qw & \targ{} & \gate{R_y(\theta_7)} & \qw 
% \end{quantikz} \\
% \caption{The hardware efficient RY-ansatz for 4 qubits with d layers.}
% \label{fig:hea}
% \end{figure}

\section{Severe dependence on initial state}
\begin{figure}[h!]
    \centering
    \includegraphics[width=.4\linewidth]{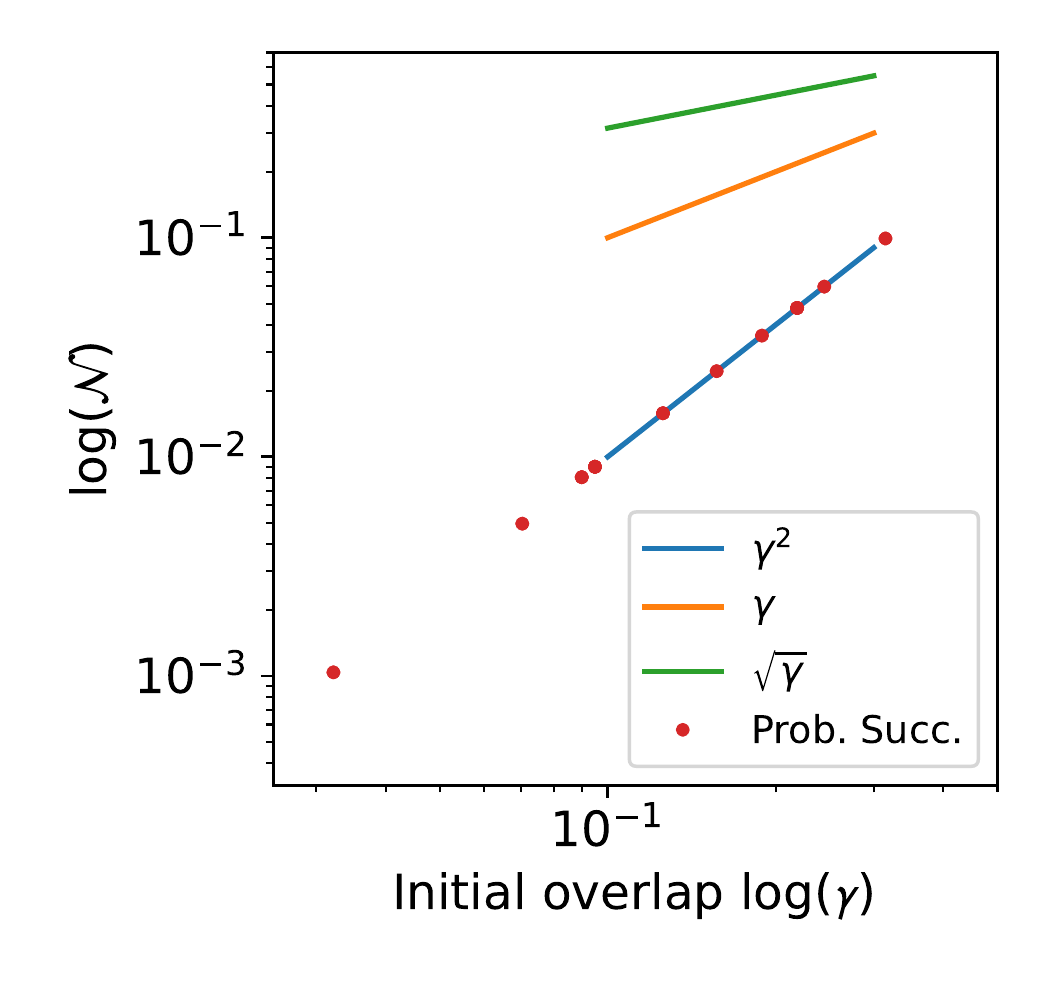}%
    \includegraphics[width=.25\linewidth]{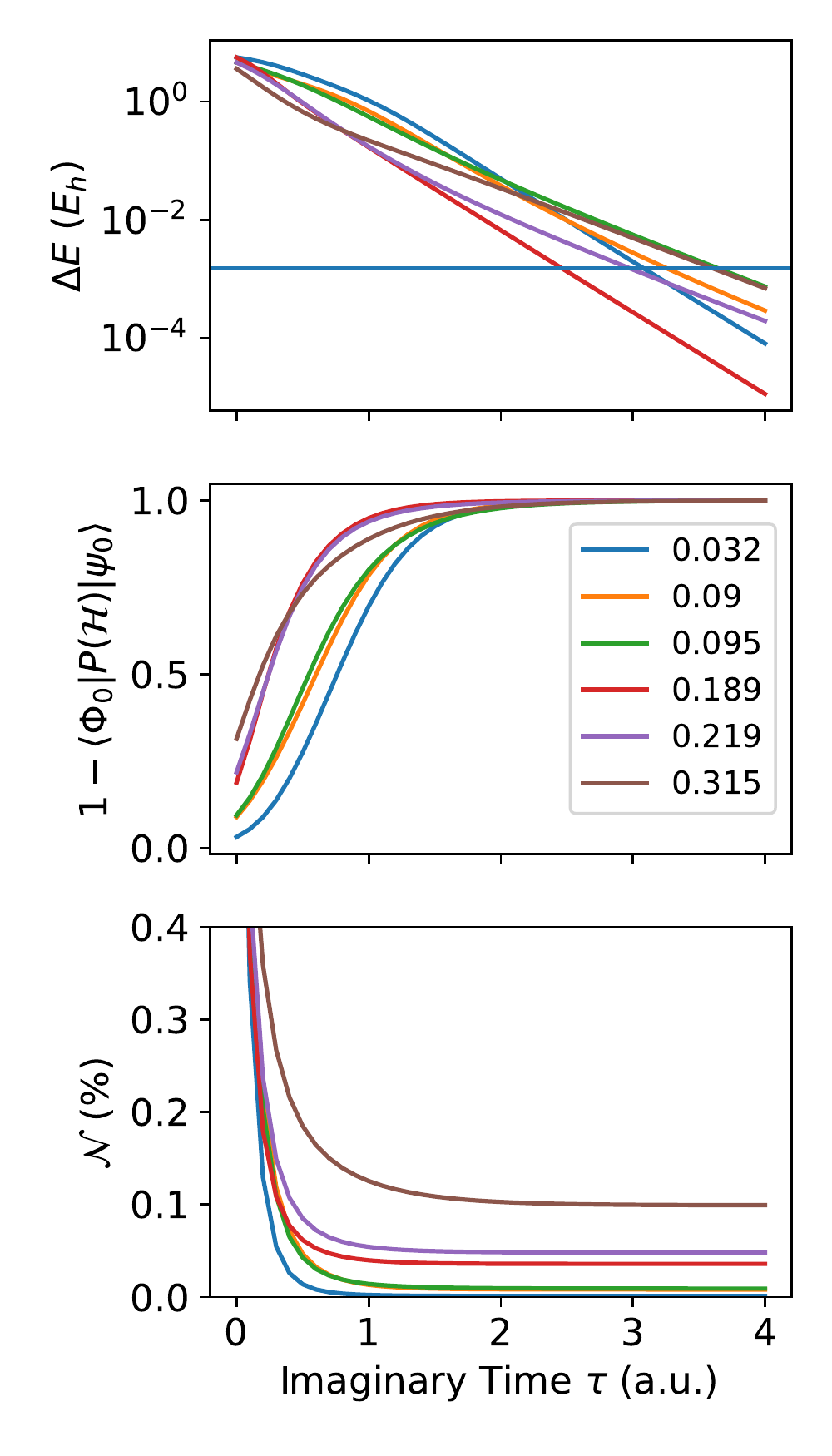}%
    \includegraphics[width=.25\linewidth]{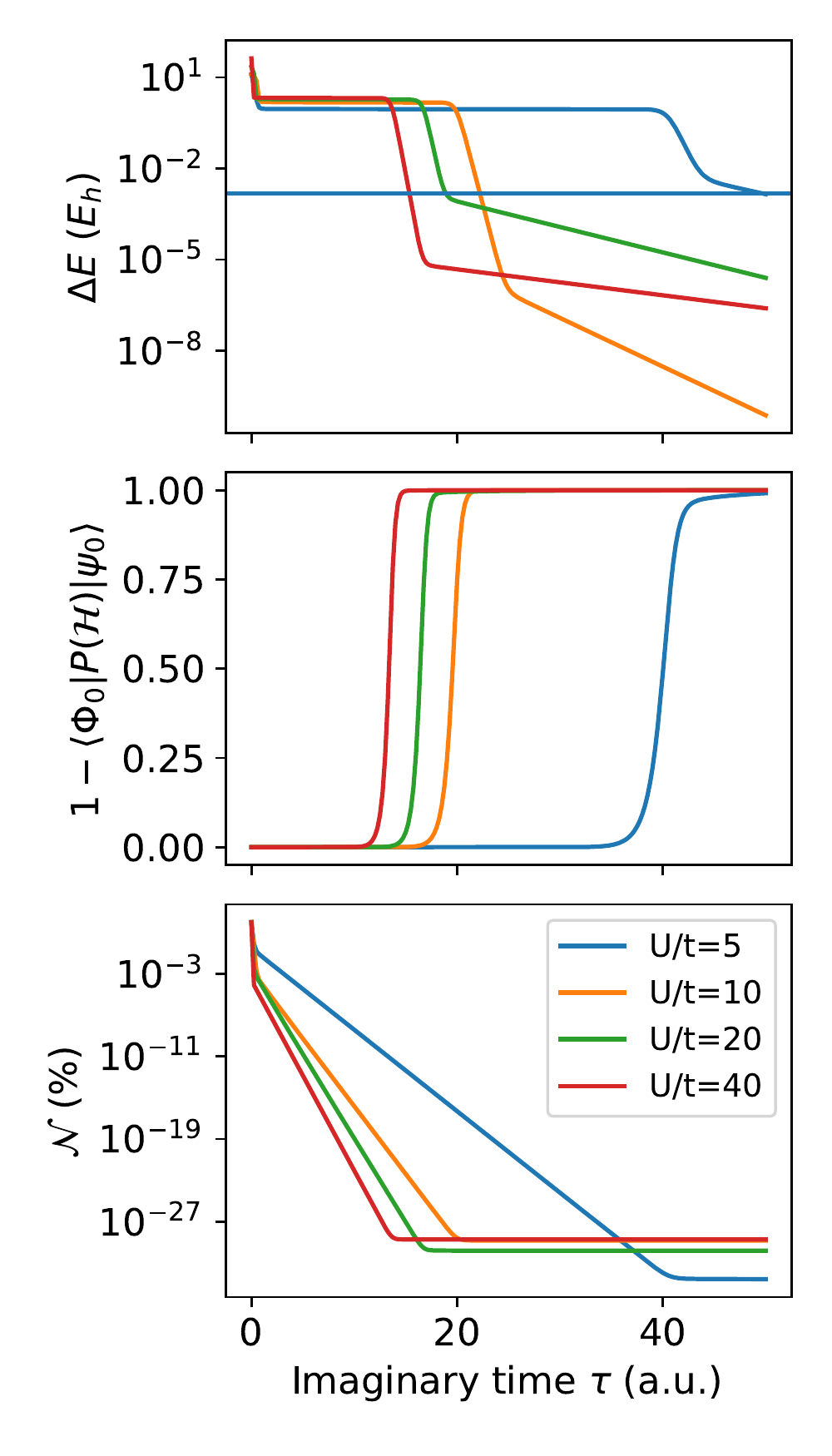}
    \caption{Probability of success and different initial trial states in the 4-site linear Fermi-Hubbard model. LEFT: Imaginary time evolution of the $U=t$ Hamiltonian with initial states of different overlaps. CENTRE: Convergence time of the $U=t$ Hamiltonian with initial states of different overlaps. RIGHT: Imaginary time evolution of different $U/t$ ratios. Horizontal lines represent chemical accuracy.} \label{fig:prob succ scale}
\end{figure}

Our analysis confirms that the probability of success depends directly on the square of the overlap between the initial state and the projected state (leftmost panel of~\cref{fig:prob succ scale}). In the case of the Fermi-Hubbard model, the strong correlation regime is notorious for the poor overlap between single reference determinants and the true ground state. This is apparent when we look at the imaginary time needed to converge Hubbard models in the strong correlation regime, where $U>>t$ (rightmost column of panels in~\cref{fig:prob succ scale}). In particular, the expected probability of a successful state preparation (calculated from if we could exactly prepare the transform $e^{-i\mathcal{H}}$) can become prohibitively small.
\end{document}